\documentclass[11pt]{article}

\usepackage{tikz}
\usepackage{graphicx,wrapfig,lipsum}   

\usepackage{xspace}

\usepackage{amsmath,amssymb}
\usepackage{amsthm}
\usepackage{mathtools}
\usepackage[noend]{algorithmic}
\usepackage[ruled,vlined]{algorithm2e}
\usepackage{url}
\usepackage{fullpage}
\usepackage{makeidx}
\usepackage{enumerate}
\usepackage[top=1in, bottom=1.25in, left=1in, right=1in]{geometry}
\usepackage{graphicx,float,psfrag,epsfig,caption}

\usepackage{hyperref}
\usepackage{epstopdf}
\usepackage{color}
\usepackage{xr}
\usepackage{subfig}
\usepackage{caption}
\usepackage[utf8]{inputenc}

\usepackage{scalerel,stackengine}

\usepackage{enumitem}

\usetikzlibrary{trees}

\stackMath
\newcommand\reallywidehat[1]{%
\savestack{\tmpbox}{\stretchto{%
  \scaleto{%
    \scalerel*[\widthof{\ensuremath{#1}}]{\kern.1pt\mathchar"0362\kern.1pt}%
    {\rule{0ex}{\textheight}}
  }{\textheight}%
}{2.4ex}}%
\stackon[-6.9pt]{#1}{\tmpbox}%
}
\usepackage[mathscr]{euscript}

\DeclareSymbolFont{rsfs}{U}{rsfs}{m}{n}
\DeclareSymbolFontAlphabet{\mathscrsfs}{rsfs}

\numberwithin{equation}{section}

\newtheoremstyle{myexample} 
    {\topsep}                    
    {\topsep}                    
    {\rm }                   
    {}                           
    {\bf }                   
    {.}                          
    {.5em}                       
    {}  

\newtheoremstyle{myremark} 
    {\topsep}                    
    {\topsep}                    
    {\rm}                        
    {}                           
    {\bf}                        
    {.}                          
    {.5em}                       
    {}  

\theoremstyle{myremark}

\theoremstyle{myremark}

\theoremstyle{myexample}

\definecolor{darkgreen}{rgb}{0.0, 0.5, 0.0}

\newcommand{\bea}{\begin{eqnarray}}
\newcommand{\eea}{\end{eqnarray}}
\newcommand{\<}{\langle}
\renewcommand{\>}{\rangle}
\newcommand{\E}{{\mathbb E}}

\def\sTV{\mbox{\tiny \rm TV}}

\def\salg{\mbox{\tiny \rm alg}}
\def\sTAP{\mbox{\tiny \rm TAP}}

\def\Unif{{\sf Unif}}

\def\eps{{\varepsilon}}

\def\id{{\boldsymbol{I}}}

\def\supp{{\rm supp}}

\def\cuF{\mathscrsfs{F}}
\def\cuE{\mathscrsfs{E}}
\def\cuU{\mathscrsfs{U}}
\def\cuL{\mathscrsfs{L}}

\def\btau{{\boldsymbol{\tau}}}
\def\bsigma{{\boldsymbol{\sigma}}}

\def\bm{{\boldsymbol m}}

\def\sd{{\sf d}}

\def\bg{{\boldsymbol{g}}}

\def\bzero{{\mathbf 0}}

\def\opt{\mbox{\tiny\rm opt}}

\def\reals{{\mathbb R}}

\def\normal{{\sf N}}

\def\bz{{\boldsymbol{z}}}
\def\bx{{\boldsymbol{x}}}

\def\bH{\boldsymbol{H}}

\def\hbz{\hat{\boldsymbol{z}}}

\def\Par{{\sf P}}
\def\de{{\rm d}}

\def\prob{{\mathbb P}}
\def\E{{\mathbb E}}

\def\<{\langle}
\def\>{\rangle}

\def\sign{{\rm sign}}

\def\cN{{\cal N}}

\def\cQ{{\cal Q}}

\def\by{{\boldsymbol{y}}}

\def\S{{\mathbb S}}

\def\para{{\boldsymbol \pi}}
\def\he{\hat{e}}
\def\2spin{\mbox{\tiny\rm 2spin}}
\def\3spin{\mbox{\tiny\rm 3spin}}

\def\cool{\mbox{\rm\tiny cool}}
\def\Glauber{\mbox{\tiny \rm Glauber}}
\def\alg{\mbox{\tiny \rm alg}}
\def\th{\mbox{\tiny \rm th}}

\def\b0{{\boldsymbol{0}}}

\def\bfone{{\boldsymbol 1}}

\def\bF{{\boldsymbol F}}
\def\bG{{\boldsymbol G}}

\DeclareMathOperator*{\plim}{p-lim}

\def\cS{{\mathcal S}}

\def\br{{\boldsymbol r}}

\newcommand{\one}{\mathbf{1}}

\newcommand{\x}{{\boldsymbol x}}
\newcommand{\y}{{\boldsymbol y}}
\newcommand{\z}{{\boldsymbol z}}
\newcommand{\w}{{\boldsymbol w}}

\renewcommand{\u}{{\boldsymbol u}}

\renewcommand{\b}{{\boldsymbol b}}

\newcommand{\R}{\mathbb{R}}
\newcommand{\T}{\mathbb{T}}

\def\hbz{\hat{\boldsymbol z}}

\def\bfone{{\boldsymbol 1}}

\newcommand{\ns}{n_{\texttt{s}}}
\newcommand{\nq}{n_{\texttt{q}}}
\newcommand{\nx}{n_{\texttt{x}}}
\newcommand{\xmax}{\texttt{x}_{\texttt{max}}}

\newcommand{\proj}{\textup{Proj}}

\title{Algorithmic Thresholds in Mean Field Spin Glasses}

\author{Ahmed El Alaoui\thanks{Simons Institute for the Theory of Computing, UC Berkeley.}\;\;\;\; and\;\;\; Andrea Montanari\thanks{Department of Electrical Engineering and
  Department of Statistics, Stanford University.}}

\date{}

\begin{document}

\maketitle

\begin{abstract}
  Optimizing a high-dimensional non-convex function is, in general, computationally hard and many
  problems of this type are hard to solve even approximately. Complexity theory characterizes
  the optimal approximation ratios achievable in polynomial time in the worst case.
  On the other hand, when the objective function is random, worst case approximation ratios are overly pessimistic.
  Mean field spin glasses are canonical families of random energy functions over the discrete hypercube $\{-1,+1\}^N$.
  The near-optima of these energy landscapes are organized according to an ultrametric tree-like structure, which
  enjoys a high degree of universality. Recently, a precise connection has begun to  emerge between
  this ultrametric structure and the optimal approximation ratio achievable in polynomial time in the typical case.
  A new approximate message passing (AMP) algorithm has
  been proposed that leverages this connection. The asymptotic behavior of this algorithm has been analyzed, conditional on the
  nature of the solution of a certain variational problem.
  
  In this paper we describe the first implementation of this algorithm and the first numerical solution of the associated variational problem.
  We test our approach on two prototypical mean-field spin glasses: the Sherrington-Kirkpatrick (SK) model, and the $3$-spin
  Ising spin glass.
  We observe that the algorithm works well already at moderate sizes ($N\gtrsim 1000$) and its behavior is consistent with
  theoretical expectations. For the SK model it asymptotically achieves arbitrarily good approximations of the global optimum.
  For the $3$-spin model, it achieves a constant approximation ratio that is predicted by the theory, and it appears to beat the
  `threshold energy' achieved by Glauber dynamics.
  Finally, we observe numerically that the intermediate states generated by the algorithm have the properties of ancestor states in the ultrametric tree. 
\end{abstract}

\section{Introduction and background}
\label{sec:Introduction}

\subsection{Mean field spin glasses}

Mean field spin glasses were introduced nearly half a century ago as idealized mathematical
models for disordered magnetic materials \cite{sherrington1975solvable}. 
Since then, they have emerged as canonical models for random
energy functions in high dimension \cite{TalagrandBook}. In this paper we will focus on the most classical of these models,
the mixed $p$-spin Ising spin glass model. This can be defined as the energy function on the discrete hypercube
$H_N:\{-1,+1\}^N\to\reals$ such that $(H_N(\bsigma))_{\bsigma\in\{-1,+1\}^N}$ is a centered Gaussian vector
with $\E\{H_N(\bsigma)H_N(\btau)\} = N\xi(\<\bsigma,\btau\>/N)$ for a
fixed analytic function  $\xi:[-1,1]\to \reals$.
Such a process can be constructed explicitly as
\begin{align}\label{eq:hamiltonian}
  H_N(\bsigma) &=  \sum_{k=2}^{\infty} \frac{c_k}{N^{(k-1)/2}}\sum_{1\le i_1 ,\cdots , i_k\le N} G_{i_1,\cdots,i_k}^{(k)}\sigma_{i_1}\cdots\sigma_{i_k}\, ,
\end{align}
where $(G^{(k)}_{i_1,\dots,i_k})_{k\ge 2, i_1,\dots, i_k\ge 1}$ is a collection of independent
standard normal random variables $G^{(k)}_{i_1,\dots,i_k}\sim\normal(0,1)$.
The two definitions are connected via the relation $\xi(x) = \sum_{k\ge 2}c_k^2 x^k$. In the following we will also occasionally
refer to the `spherical' version of  this model, in which the constraint $\bsigma\in\{-1,+1\}^N$ is replaced by
$\bsigma\in\S^{N-1}(\sqrt{N})$ (the sphere of radius $\sqrt{N}$ in $\reals^N$).

We will be concerned with the superlevel sets of this Hamiltonian, namely sets
$\cS_N(\eps)$  of configurations $\bsigma\in\{-1,+1\}^N$ such that $H_N(\bsigma) \ge (1-\eps) \max_{\bsigma'\in\{\pm 1\}^N}
H_N(\bsigma')$. We will ask whether configurations in $\cS_N(\eps)$ can be found in polynomial time for any constant $\eps>0$.
From a physics point of view, superlevel sets (with small $\eps$) determine the the low-temperature behavior of the system
that is described by  the Gibbs measure $\mu_{N,\beta}(\bsigma) := \exp\{\beta H_N(\bsigma)\}/Z_{N,\beta}$.

\subsection{Geometry of near-optima}

Within a decade from their introduction,  physicists unveiled a beautiful probabilistic structure
that captures the organization of these near-optima \cite{SpinGlass}. 
This structure is referred to as `replica symmetry breaking' (RSB)  in connection to its first discovery via the
so-called replica method. It is useful to summarize the main elements of this picture, as first outlined in \cite{mezard1984nature,mezard1985microstructure}.

\begin{figure}[t]
  \phantom{A}\hspace{1cm}
  \begin{tikzpicture}[thick, level distance=8mm, inner sep = 0.2mm]
    \draw[very thick, ->] (-4,0) -- (-4,-4);
    \node at (-3.75,-2) [rotate = 90] {$t=q$};
    \node at (1,-1.7) {$\gamma$};
    \node at (0.3,-4.2) {$\alpha_1$};
    \node at (1.35,-4.2) {$\alpha_2$};
  \tikzstyle{level 1}=[sibling distance=30mm]
  \tikzstyle{level 2}=[sibling distance=15mm]
  \tikzstyle{level 3}=[sibling distance=7.5mm]
  \tikzstyle{level 4}=[sibling distance=3.75mm]
  \tikzstyle{level 5}=[sibling distance=1.875mm]
   \node at (0,0) [circle,draw=black!100,fill=black!100] {}
    child foreach \x in {0,1}
      {child foreach \y in {0,1}
        {child foreach \z in {0,1}
          {child foreach \w in {0,1}
            {child foreach \u in {0,1}}} }};
  \tikzstyle{level 1}=[sibling distance=20mm, level distance=22mm]
  \tikzstyle{level 2}=[sibling distance=7.5mm, level distance=6mm]
  \tikzstyle{level 3}=[sibling distance=3.75mm, level distance=6mm]
  \tikzstyle{level 4}=[sibling distance=1.875mm, level distance=6mm]
  \node at (7,0) [circle,draw=black!100,fill=black!100] {}
     child foreach \x in {0,1,2}
      {child foreach \y in {0,1}
        {child foreach \z in {0,1}
          {child foreach \w in {0,1}} }};
      \draw [dashed] (4,0) -- (10,0);
      \draw [dashed] (4,-2.2) -- (10,-2.2);
      \draw [dashed] (4,-4) -- (10,-4);
      \node at (9.7,-0.3) {$q=0$};
      \node at (9.7,-2.5) {$q_0$};
      \node at (9.7,-4.3) {$q_*$};
    \end{tikzpicture}
    \caption{Cartoon of the tree of ancestor states: two examples. Tree levels are indexed by the overlap value $q$,
      which corresponds to the time index in the algorithm evolution $t$, and to the norm of magnetization vectors
      $\|\bm^{\gamma}\|_2^2/N=q$ (if $\gamma$ is at level $q$). On the right: a tree with large overlap gap between $0$ and $q_0$.}
    \label{fig:TreeCartoon}
  \end{figure}
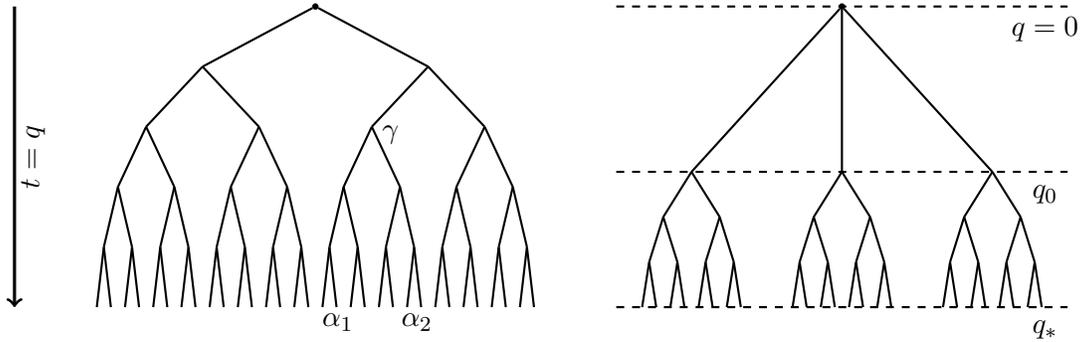
  
For  $\eps$ a sufficiently small constant, the random set $\cS_N(\eps)$ can be partitioned into $M$ `pure states'
$\cS_{N}^{\alpha}$, $1\le \alpha\le M$, plus eventually a negligible  subset of configurations $\cN_N$:
$\cS_N(\eps)=\cup_{\alpha=1}^m\cS_{N}^{\alpha}\cup \cN_N$.
Each pure state $\cS_N^{\alpha}$ can be characterized by its barycenter $\bm^{\alpha}$ (the `magnetization' vector).
All of these barycenters approximately lie on a sphere of radius $\sqrt{Nq_*(\eps)}$: $\|\bm^{\alpha}\|_2^2 = Nq_*(\eps) (1+o_N(1))$.
The rescaled radius $q_*$ is referred to as the Edwards-Anderson parameter.
Within a pure state, the distance from the barycenter concentrates, which implies $\|\bsigma-\bm^{\alpha}\|^2_2= N(1-q_*(\eps))(1+o_N(1))$ for almost all $\bsigma\in \cS_N^{\alpha}$ (we can redefine $\cN_N$ to include a small
number of configurations that violate this condition). 
Pure states are organized in a tree whose levels can be
indexed by $q\in \cQ\subseteq [0,q_*]$ as we describe next. Pure states occupy the leaves at level $q_*$.
Each internal node $\gamma$ in this tree is referred to as an `ancestor state', and can be
associated to a set of configurations $\cS^{\gamma}_N$ that is formed by the union of pure
states that are its descendants: $\cS^{N}_{\gamma}:=\cup_{\alpha\in D(\gamma)}\cS^{\alpha}_N$ (here $D(\gamma)$ is the set of leaves that are
descendants of $\gamma$).
By averaging over these states, we can associate to the ancestor  $\gamma$ a barycenter $\bm^{\gamma}$: the level $q$
of node $\gamma$ corresponds to the norm of this barycenter, $\|\bm^{\gamma}\|_2^2= Nq(1+o_N(1))$.

Given a certain  ancestor state $\gamma$ at level $q_1$, the radius $\|\bm^{\alpha}-\bm^{\gamma}\|_2$ concentrates as $\alpha$ varies among nodes at level $q_2>q_1$ that are descendants of  $\gamma$. Note that
$\sum_{\alpha}w_{\alpha}\<\bm^{\alpha}-\bm^{\gamma},\bm^{\gamma}\>=0$
by  construction (where $w_{\alpha} :=|\cS_N^{\alpha}|/|\cS_N^{\gamma}|$). Hence we must have
$\|\bm^{\alpha}-\bm^{\gamma}\|_2^2=N(q_2-q_1+o_N(1))$ and
$\<\bm^{\alpha}-\bm^{\gamma},\bm^{\gamma} \> = o_N(N)$. A cartoon of the geometry of pure states and ancestor states
is given in Fig.~\ref{fig:TreeCartoon}. Each internal node $\gamma$ corresponds
to an ancestor state and its descendant lie close to the hyperplane
orthogonal to $\bm^{\gamma}$. As a special consequence of this geometric picture, given two pure states $\alpha_1$,
$\alpha_1$, with closest common ancestor $\gamma$, we have $\<\bm^{\alpha_1},\bm^{\alpha_2}\> =
\|\bm^{\gamma}\|_2^2(1+o_N(1))$. This means that we can directly probe the tree structure of pure states by sampling two independent configurations $\bsigma^1,\bsigma^2\sim_{iid}\Unif(\cS_N(\eps))$. The expected (over the realization of $H_N)$
probability distribution of the `overlap' $|\<\bsigma^1,\bsigma^2\>|/N$ converges to a limit measure $\nu_{\eps}$ with support
$\cQ\subseteq [0,1]$.

In over the thirty years since the RSB picture was first described, elements of it have emerged in an impressive variety of models
from random constraint satisfaction problems, where it describes the geometry of the set of solutions, to random combinatorial
optimization, where it describes the structure of near-optima,
to communications and statistical estimation where it applies to near optimal reconstructions. We
refer to \cite{engel2001statistical,NishimoriBook,MezardMontanari} for overviews of these research areas.

In the context of mixed $p$-spin models, the ultrametric structure of pure states has been made rigorous
in a remarkable sequence of mathematical works \cite{panchenko2013parisi,panchenko2013sherrington,chen2018generalized,chen2019generalized}. Among the wealth of results, we know that at sufficiently low temperatures
(small $\eps$ or large $\beta$) the number of levels of the tree (i.e.\ the ${\rm card}(\cQ)$)
becomes arbitrarily large \cite{auffinger2020sk}.
On the other hand, given a sequence of coefficients $(c_k)_{k\ge 2}$ (or, equivalently, their generating
function $\xi(x)$), we do not yet know what is the qualitative structure of the set $\cQ$ which indexes the levels of the tree,
and the possible values of the overlap between pure states.

\subsection{Algorithms?}

Does the universal RSB structure of random energy landscapes have algorithmic consequences?
The present paper focuses on this question, and more precisely on the
\emph{search problem}.
Namely, we are looking for a polynomial-time  algorithm that takes as input a specification of
the Hamiltonian $H_N(\,\cdot\,)$, and returns as output a vector $\bsigma^{\salg}\in\{-1,+1\}^N$
such that
\begin{align}
  \lim_{N\to\infty}\prob\Big(H_N(\bsigma^{\salg})\ge (1-\eps)\max_{\bsigma\in\{\pm 1\}^N}H_N(\bsigma)\Big) =1\, .
  \label{eq:Prob}
  \end{align}
  In particular, we are interested in understanding which of these two scenarios holds:
  \begin{itemize}
\item[$(i)$] An arbitrarily good approximation of $\max_{\bsigma\in\{\pm 1\}^N}H_N(\bsigma)$
  can be achieved in polynomial time, i.e.\ the above holds for any fixed $\eps>0$; 
  \item[$(ii)$] Only a constant factor approximation can be achieved,
  i.e.\ the above holds only for $\eps>\eps_*$ for some strictly positive $\eps_*$. In the latter case we want to determine $\eps_*$.
  \end{itemize}

  The relevance of spin glass theory for optimization with random structures was
  noted early on in the history of the subject. Quoting from a seminal paper by M\'ezard and Virasoro \cite{mezard1985microstructure}:
\begin{quote}
  `The ultrametric topology of the space of equilibrium
states of a spin glass deserves special attention. Such an
organization might exist in other systems with frustration
and disorder, and it should have consequences in
such fields as optimization problems or neural
networks.'
\end{quote}
Indeed, the connection between statistical physics and optimization has been an object
of interest at least since the introduction of simulated annealing \cite{kirkpatrick1983optimization}. Beginning in the eighties,
physicists used nonrigorous methods from physics, such as the replica method (or the equivalent cavity method),
to study the typical properties
of random combinatorial optimization problems. Notable examples of this literature include the
assignment problem \cite{mezard1985replicas}, the random traveling salesman problem \cite{mezard1986replica}, random
$K$-satisfiability ($K$-SAT) \cite{monasson1999determining,mezard2002analytic,krzakala2007gibbs}, and coloring random graphs \cite{mulet2002coloring}.

The idea of using the replica or cavity methods as algorithmic tools was only developed more
recently, beginning with the work of M\'ezard, Parisi and Zecchina on random $K$-SAT \cite{mezard2002analytic}.
These authors proposed a message passing algorithm known as `survey propagation' that amounts to iteratively
solving the `one step RSB' cavity equation for a given $K$-SAT formula. The solution of these equations is then used to
iteratively assign the values of variables. Empirical studies demonstrated that this approach is effective in finding solutions of
large random $3$-SAT formulae \cite{braunstein2005survey}. On the other hand, rigorous analysis was carried out for large $K$,
suggesting limited advantage offered by survey propagation over simpler heuristics \cite{gamarnik2017performance}.

It should be emphasized that, unlike the classical $p$-spin model of Eq.~\eqref{eq:hamiltonian},
the $K$-SAT model can be regarded as a spin glass on a \emph{sparse} random graph, a
setting which presents additional technical challenges. 

Among  negative results, Gamarnik and coauthors established ---in several examples--- a connection between
the support $\cQ$ of the overlap distribution and the ability of certain classes of algorithms to approximately solve the
underlying optimization problem \cite{gamarnik2014limits,chen2019suboptimality}. Namely, if $\cQ$ is not an interval, then local algorithms
fail: the class of local algorithms include certain implementations of survey propagation and other message passing techniques.
The condition of $\cQ$ not being an interval is also referred to as `overlap gap condition'.

The idea of exploiting the RSB picture of the energy landscape in an optimization algorithm was
revived by a sequence of papers over the last two years. 
Addario-Berry and Maillard \cite{addario2018algorithmic} study optimization of an idealized model for the tree of
ancestor states in spin glasses, also known as the Generalized Random Energy Model (or GREM) \cite{derrida1985generalization}.
They consider optimization algorithms that descend the tree with limited lookahead and prove that such algorithms can find a near optimum efficiently, if and only if the tree
satisfies the no-overlap gap condition. Subag \cite{subag2018following} studied the general spherical $p$-spin model and presented
an algorithm that ---again--- returns a near optimum under the no-overlap gap condition. More
generally, \cite{subag2018following}   gives an explicit formula for the energy achieved by this algorithm. Unlike for the Ising spin glass,
for the spherical model it is explicitly known which generating functions $\xi$ imply no
overlap gap.

Finally, in \cite{montanari2019optimization,alaoui2020optimization}, the present authors and
Mark Sellke developed a nearly linear time approximate message passing (AMP)  algorithm
for the Ising spin glass and characterized the value it achieves. This characterization
holds \emph{under the assumption} that a certain variational problem
over functions on the interval $[0,1]$ achieves its infimum. As a special case, whenever the generating function
$\xi$ is such that the resulting model has no overlap gap, this algorithm achieves \eqref{eq:Prob}
for any $\eps>0$. 

This paper presents the following contributions:
\begin{description}
\item[Variational principle.] As mentioned above, the value achieved asymptotically by the algorithm of
  \cite{montanari2019optimization,alaoui2020optimization} is described by a variational principle that can be viewed as a modification of the celebrated Parisi formula.
  Is the infimum of this variational principle achieved? Can it be evaluated numerically?
  We describe a procedure for solving the variational principle, and apply it to two examples, 
  showing empirically that we obtain approximate solutions with the desired properties.
\item[Implementation.] The analysis of \cite{montanari2019optimization,alaoui2020optimization} is asymptotic and establishes guarantees of the form
  \eqref{eq:Prob}, that hold as $N\to\infty$. How large should $N$ be for these asymptotics to kick in?
  Do these algorithms also work reasonably well for ---say--- $N$ the order of $10^3$?
  Finally, for any given $N$, several implementation choices need to be made, which become irrelevant as $N\to\infty$.
  We propose a specific implementation of the algorithm, which appears to perform well already for $N$ of the order of
  a few thousands, and study the leading finite-$N$ effects.
\item[Threshold energy.] We study two specific examples (two choices of the generating function $\xi$):
  the Sherrington-Kirkpatrick model and the 3-spin model. In the second case the algorithm only achieves a
  threshold energy that is  smaller  ---by a constant factor--- than the global optimum. We compare this threshold
  with the energy achieved by Glauber dynamics (at fixed very low temperature) and show that the new algorithm
  achieves a better approximation.
\item[Connection with physics.] In what ways does the algorithm of \cite{alaoui2020optimization} explore the landscape of
  ancestor states? We elucidate this question in two different ways. First we show that the path followed by the algorithm during its execution is such that the configuration $\bm^t$ produced at time $t\in[0,1]$ has the same properties
  as an ancestor state at level $t$. In particular, it is an approximate solution of the generalized TAP equations and has a
  nearly optimal value of the generalized TAP free energy \cite{mezard1985microstructure,chen2018generalized,chen2019generalized}.

  Second, we show that the algorithm can be modified
  as to produce not one, but multiple near optima $(\bsigma^\alpha)_{\alpha\ge 1}$, with
  $\<\bsigma^{\alpha},\bsigma^{\gamma}\>/N\approx 0$ for $\alpha\neq \gamma$.
\end{description}

\section{Description of the algorithm}
\label{sec:Description}

The algorithm of \cite{montanari2019optimization,alaoui2020optimization} is iterative. It is convenient to index iterations (time) by  $\T_{\delta}:=\{0,\delta,2\delta,\dots, 1-\delta,1\}$,
where $\delta$ is a step size that should be though of as vanishing when $N\to\infty$, but sufficiently slowly. The algorithm generates
a sequence $\bm^t\in \R^N$, $t\in\T_{\delta}$, where each vector $\bm^t$ is a function of
$(\bm^s)_{s\in\T_{\delta}\cap[0,t-\delta]}$. The interpretation of these quantities is as follows: at time $t$, the algorithm has effectively
selected a subset $\cS_{N,t}\subseteq\{-1,+1\}^N$ of near optima. The vector $\bm^t$ should be interpreted as the
barycenter of $\cS_{N,t}$ (or magnetization vector), and $t$ the corresponding overlap. In particular, we should expect
$\bm^t\in[-1,1]^N$, $\|\bm^{t}\|^2_2\approx Nt$, and $\|\bm^{t'}-\bm^t\|^2_2\approx N(t'-t)$ for all  $t'>t$.
We will see that these conditions are indeed verified as $N\to\infty$.

Before getting into a detailed description of the iteration that computes the sequence of vectors $\bm^t$,
it can be useful to give a heuristic discussion of some of its elements. At time $t$ we want to select a subset
$\cS_{N,t}\subseteq \cS_{N,t-\delta}$ of near optima, given that we know the current barycenter $\bm^{t-\delta}$. It
would seem reasonable to linearize the objective function around the barycenter and compute the current gradient:
$\hbz^{t} = \nabla H_N(\bm^{t-\delta})$. We could then try to compute the new barycenter as the expectation of the `linearized'
Gibbs measure $e^{\beta\<\hbz^t,\bsigma\>}/Z_t$. This would result into a second equation $\bm^t= \tanh(\beta \hbz^t)$.

The actual algorithm is more complex,
although it retains some of the structure suggested by the above naive argument. It is a special example of approximate message passing (AMP) algorithm \cite{DMM09,BM-MPCS-2011}, and we will refer to it as IAMP (incremental AMP).
We introduce two auxiliary sequences $(\bx^t)_{t\in\T_{\delta}}$, $(\bz^t)_{t\in\T_{\delta}}$, and define $\de\bm^t :=\bm^{t+\delta}-\bm^t$
(and $\de\bx^t$, $\de\bz^t$ analogously). Starting from $\bz^0=\bm^0=\bx^0=\bzero$,
and $\bz^{\delta}\sim\normal(0,\delta\id_N)$, we compute recursively 
\begin{align}
  \de\bz^t & = \nabla H_N(\bm^t) -\nabla H_N(\bm^{t-\delta}) +\hat{d}(t)\de\bm^{t-\delta} \, ,\label{eq:ALG1}\\
  \de\bx^t &= \xi''(t)\gamma_*(t)\partial_x\Phi_{\gamma_*}(t;\bx^t)\delta +\de\bz^t\,  ,\label{eq:ALG2}\\
  \bm^t &= \partial_x\Phi_{\gamma_*}(t;\bx^t)\, , \label{eq:ALG3}
\end{align}
where $\hat{d}(t) = \xi''(t)\int_t^1\gamma(s)\, \de s$.
A feasible configuration $\bsigma^{\alg}$ is obtained by rounding $\bm^1$, e.g.\ via $\bsigma^{\alg} :=\sign(\bm^1)$,
Here $\Phi_{\gamma_*}:[0,1]\times\reals\to\reals$ and $\gamma_*:[0,1)\to\reals$ are functions which we will define next via a
variational principle
(and $\partial_x\Phi_{\gamma_*}(t;\bx^t)$ denotes the function $\partial_x\Phi_{\gamma_*}(t;\,\cdot\,)$ applied entrywise to the vector $\bx^t$.)

Given a function $\gamma:[0,1]\to \reals_{\ge 0}$, consider the following partial differential equation,
which we refer to as the Parisi PDE:
\begin{align}\label{eq:PDEFirst}
\begin{split}
\partial_t \Phi_{\gamma}(t,x)+\frac{1}{2}\xi''(t) \Big(\partial_x^2\Phi_{\gamma}(t,x)+\gamma(t) (\partial_x\Phi_{\gamma}(t,x))^2\Big) = 0, ~~~ (t,x)\in [0,1)\times \reals \, ,
\end{split}
\end{align}
with boundary condition $\Phi_{\gamma}(1,x) = |x|$. It is known that a weak solution to the above PDE exists and is unique if $\gamma\in\cuL$ \cite{jagannath2016dynamic,alaoui2020optimization}, where we define the following space of functions
\begin{align}
\cuL := \Big\{\gamma:[0,1)\to \reals_{\ge 0}: \;\; \|\xi''\gamma\|_{\sTV[0,t]}<\infty~ \forall t\in [0,1), \int_0^1\!\xi''\gamma(t)\,\de t < \infty\Big\} \, .\label{eq:LDef}
\end{align}
Here $\xi''\gamma(t):=\xi''(t)\gamma(t)$, and $ \|\xi''\gamma\|_{\sTV[0,t]}$ is the total variation of this function over the interval $[0,t]$. In particular, the condition $\|\xi''\gamma\|_{\sTV[0,t]}<\infty$ is satisfied if $\xi''\gamma$ has a finite number of jump discontinuities over the interval $[0,t]$ and has no singularities.

We use
the solution of this PDE to define the following variational problem
\begin{align}
  \Par(\gamma) &:= \Phi_{\gamma}(0,0)-\frac{1}{2}\int_{0}^1 t\xi''(t)\gamma(t)\, \de t\, ,\label{eq:ParisiFunctional}\\
  e_{\alg}&:= \inf_{\gamma\in\cuL} \Par(\gamma)\, . \label{eq:VarPrinciple}
\end{align}
If the infimum above is achieved at $\gamma_*\in\cuL$, using this function in the construction of the algorithm of Eqs.~\eqref{eq:ALG1}
to \eqref{eq:ALG3} yields $H_N(\bsigma^{\alg})/N\to e_{\alg}$ \cite{alaoui2020optimization}. In other words, the variational principle of
Eq.~\eqref{eq:VarPrinciple} characterizes the function value achieved by the IAMP algorithms\footnote{The algorithm
  analyzed in \cite{alaoui2020optimization} differs from the one of Eqs.~\eqref{eq:ALG1} to \eqref{eq:ALG3} by terms that vanish as
  $N\to\infty$, $\delta\to 0$. See below for further discussion.}.

The variational principle \eqref{eq:VarPrinciple}  is intimately related to the celebrated
Parisi formula for the optimum value. Namely, we define
\begin{equation}\label{eq:original}
  e_{\opt}: =\inf_{\gamma\in \cuU}\Par(\gamma), \;\;\;\;\;\;\;
  \cuU \equiv \cuL \cap \{\gamma~\mbox{non-decreasing}\}
\end{equation}
Then $\lim_{N\to\infty}\max_{\bsigma\in\{\pm 1\}^N}H_N(\bsigma)/N = e_{\opt}$
\cite{talagrand2006parisi,panchenko2013sherrington,auffinger2017parisi}. Notice that, since $\cuU\subseteq\cuL$,
we have $e_{\opt}\ge e_{\alg}$ as it should be. Further, $\gamma\mapsto \Par(\gamma)$ is known to be strictly convex
\cite{auffinger2015parisi,jagannath2016dynamic}. Hence, if the infimum in Eq.~\eqref{eq:original} is achieved on 
$\gamma_*$ which is strictly increasing (no overlap gap), the  constraint $\{\gamma~\mbox{non-decreasing}\}$
is not active at $\gamma_*$. Hence $\gamma_*$ is 
also a minimizer for the variational principle \eqref{eq:VarPrinciple} and therefore $e_{\alg}= e_{\opt}$. In other words, under no overlap gap,
the above algorithm achieves a $(1-\eps)$-approximation of the optimum for any $\eps>0$ \cite{alaoui2020optimization}.

From a geometric point of view, the last term in Eq.~\eqref{eq:ALG1} is chosen in such a way that $\de\bz^{t}$ is approximately
orthogonal to $(\bm^s)_{s\le t}$, $(\bz^s)_{s\le t}$. Hence, this update can be replaced by a similar one in which
\begin{align}
  \de\tilde{\bz}^{t} &= \proj_{t-\delta}^{\perp}[\nabla H_N(\tilde{\bm}^t)-\nabla H_N(\tilde{\bm}^{t-\delta})]\, ,
\end{align}
where $\proj_{t-\delta}^{\perp}$ is the projector orthogonal to ${\rm span}(\{\tilde{\bm}^{s}\}_{s\in \T_{\delta}\cap[0,t-\delta]})$. 

The algorithm of
Eqs.~\eqref{eq:ALG1} to \eqref{eq:ALG3} is a special example of a broader class of IAMP algorithms,
and is obtained by optimizing over this class. 
In order to introduce this class, given a vector $\bx\in\reals^N$, we denote its average by
$\E_{N}(\bx):=\sum_{i=1}^Nx_i/N$. For two vectors $\bx,\by\in\reals^N$, the we also define
the normalized scalar product $\<\bx,\by\>_N:=\sum_{i=1}^Nx_iy_i/N$. The basic iteration depends on two
functions $u, v:[0,1]\times\reals\to\reals$. Starting from $\bz^0=\bm^0=\bx^0=\bzero$,
and $\bz^{\delta}\sim\normal(0,\delta\id_N)$, we proceed with the following updates 
\begin{align}
  \bz^{t+\delta} &= \nabla H_N(\bm^t)-\sum_{s\in \T_\delta\cap [\delta,t]}\sd_{t,s} \bm^{s-\delta}\, ,\label{eqs:AlgUV_1}\\
  \sd_{t,s}&:= -\xi''(\<\bm^t,\bm^{s-\delta}\>_N\big) \E_N\Big(u_{\delta}(s;\bx^{s})\bfone_{s<t}-u_{\delta}(s-\delta;\bx^{s-\delta})\Big)\, , \label{eqs:AlgUV_2}\\
  \bm^t& = \sum_{s\in \T_{\delta}\cap[0,t-\delta]} u_{\delta}(s;\bx^s)\odot (\bz^{s+\delta}-\bz^s)\, , \label{eqs:AlgUV_3}\\
  \bx^{t+\delta}&  = \bx^t+v(t;\bx^t)\delta +(\bz^{t+\delta}-\bz^t)\, , \label{eqs:AlgUV_4}\\
                  u_{\delta}(t;\bx^t) & := \frac{u(t;\bx^t)}{\E_N[u(t;\bx)^2]^{1/2}}\, . \label{eqs:AlgUV_5}
\end{align}
The algorithm of Eqs.~\eqref{eq:ALG1} to \eqref{eq:ALG3} is obtained
by optimizing the value achieved by the last procedure
over the functions $u,v$. As discussed in  \cite{alaoui2020optimization}, the variational principle \eqref{eq:VarPrinciple} is
the dual of this maximization problem.
The optimal choice of $u$ and $v$ is given as follow in terms of  the dual solution $\gamma_*$ of
\eqref{eq:VarPrinciple}:
\begin{align}
  u(t;x) & = \partial_x^2\Phi_{\gamma_*}(t;x)\, ,\;\;\;\;\;\;\; v(t;x) = \xi''(t)\gamma_*(t)\partial_x\Phi_{\gamma_*}(t;x)\, .\label{eq:OptimalUV}
\end{align}
The algorithm of Eqs.~\eqref{eq:ALG1} to \eqref{eq:ALG3} is obtained by substituting these functions in Eqs.~\eqref{eqs:AlgUV_1} to
\eqref{eqs:AlgUV_5} and noting that, almost surely, 
\begin{align*}
&\lim_{N\to\infty}\<\bm^t,\bm^{s-\delta}\>_N= s-\delta, ~~~
\lim_{\delta\to 0}\lim_{N\to\infty}\E_N[u_{\delta}(s;\bx^{s})] =\int_s^1\gamma_*(r)\, \de r,\\ 
&~~~\mbox{and}~~~\lim_{\delta\to 0}\lim_{N\to\infty} \delta^{-1}\E_N[u_{\delta}(s;\bx^{s}) -u_{\delta}(s-\delta;\bx^{s-\delta})]= -\gamma_*(s).
\end{align*}

We conclude this section by observing that the IAMP algorithm described in here admits a
generalization at non-zero temperature \cite{montanari2019optimization}.  
Instead of attempting to find a near optimum of the cost function $H_N(\bsigma)$, this
algorithm attempts to produce a magnetization vector
$\bm\in\reals^N$ corresponding to a pure state (i.e.\ a leaf in the tree of ancestor states), under a
no overlap gap condition. The algorithm proceeds again
using Eqs.~\eqref{eqs:AlgUV_1} to~\eqref{eqs:AlgUV_5} and only differs in the choice of the functions $u$, $v$.
These are chosen using Eq.~\eqref{eq:OptimalUV}, with $\Phi_{\gamma}$ the solution of the Parisi PDE \eqref{eq:PDEFirst}
with boundary condition $\Phi_{\gamma}(1,x) = \beta^{-1}\log (2\cosh\beta x)$ (instead of $\Phi_{\gamma}(1,x) = |x|$). 
Finally, $\gamma=\gamma_{\beta}$ is determined according to
\begin{align*}
\gamma_{\beta} = \arg\min\big\{\Par(\gamma)\,, 
  \mbox{subj. to} &\;\;\;\; \gamma\in\cuU\, ,\;\; \gamma(1)=  \beta \big\}\,.
\end{align*}
For models with no overlap gap, the solution $\gamma_{\beta}$ to
this problem is strictly increasing on $[0,q_*]$, and constant on $(q_*,1]$.
 The  iteration of Eqs.~\eqref{eqs:AlgUV_1} to \eqref{eqs:AlgUV_5} is halted at $t \approx  q_*$ (e.g. at $t=\delta\lfloor q_*/\delta\rfloor$).

\section{Implementation and numerical experiments}

In this section we present our implementation of the algorithm described above, and our
empirical results. We focus on the pure 2-spin and 3-spin models:
\[\xi_{\2spin}(x) = \frac{x^2}{2}, \;\;\;\;\;\;\mbox{and}\;\;\;\;\;\;\xi_{\3spin}(x) = \frac{x^3}{2}.\]
These two models are representative of qualitatively different behaviors.
The $2$-spin model, also known as Sherrington-Kirkpatrick (SK) model, was first introduced in
\cite{sherrington1975solvable} and its landcape is believed to be organized in a continuum tree with no overlap
gap: $\gamma_*$ is strictly increasing. The $3$-spin model was first introduced in \cite{derrida1980random} and subsequently studied as a toy model for structural glasses \cite{kirkpatrick1987p}.
The tree of states is believed to have a gap at $0$: ancestor state exist only for $q\in [q_0,q_*]$, $q_0>0$, and are well separated.  Given a typical pair $\gamma_1$, $\gamma_2$ of  ancestor states at level
$q_0$, we have $\|\bm^{\gamma_1}\|_2^2, \|\bm^{\gamma_2}\|_2^2=Nq_0(1+o_N(1))$ while $\<\bm^{\gamma_1},\bm^{\gamma_2}\> = o(N)$. Further, each of these ancestor states gives rise to a continuous  tree of descendants.

\subsection{Numerical solution of the variational principles}
\label{sec:variational_principle}
We compute approximate values of the variational problems Eq.~\eqref{eq:VarPrinciple}
and Eq.~\eqref{eq:original} by considering piecewise constant functions
$\gamma(t) = \sum_{i=1}^{\nq} r_i \one_{t \in [q_i,q_{i+1})}$ with
$\nq$ steps located on a uniform grid
$q_i = \frac{i-1}{\nq}$. The Parisi PDE~\eqref{eq:PDEFirst} can then be solved via the Cole-Hopf transform backwards in time:
\begin{equation}\label{eq:cole-hopf}
\Phi_{\gamma}(q_{i},x) = \frac{1}{r_i} \log \E \Big[\exp\Big(r_i \Phi_{\gamma}\big(q_{i+1},x + \sqrt{\xi'(q_{i+1}) - \xi'(q_i)}Z\big)\Big) \Big]~~~\forall 1\le i \le n,
\end{equation}        
where $Z \sim \normal(0,1)$. 
We restrict the space variable $x$ to an interval  $[-\xmax,\xmax]$ which is also discretized into a uniform grid with spacing $1/\nx$.
This discretization scheme depends on the set of parameters $\para:=(\nq, \nx,\xmax)$.
In our experiments we observe that $\nx=500$, and $\xmax=10$ is sufficient to produce negligible discretization errors,
and study the dependence on $\nq$. 

The Parisi functional becomes a convex function of the vector $\br = (r_i)_{i=1}^{\nq} \in \reals_{\ge 0}^{\nq}$,
which we denote by $\Par_{\para}(\br)$.  
The variational principles Eq.~\eqref{eq:VarPrinciple} and Eq.~\eqref{eq:original} are then approximated by the finite-dimensional convex optimization problems 
\begin{equation}\label{eq:finite}
  \he_{\alg}(\para):= \min_{\br \in \reals_{\ge 0}^{\nq}}\Par_{\para}(\br) \,,\qquad \qquad
  \he_{\opt}(\para):= \min_{\br \in \reals_{\ge 0}^{\nq}}\big\{\Par(\br)\;\ \mbox{subj. to}\;\; r_1\le \dots\le r_{\nq}\big\}\, .
\end{equation}  
The above optimization problems are solved to near-optimality via projected gradient descent with backtracking line search
\cite{parikh2014proximal}. The gradient of $\Phi_{\gamma}(0,0)$ w.r.t.\ $\br$ is computed recursively via the formula~\eqref{eq:cole-hopf}.
(We refer to Appendix \ref{sec:NumDetails}  for further details.)

\begin{figure}
\centering
\includegraphics[width=.45\textwidth]{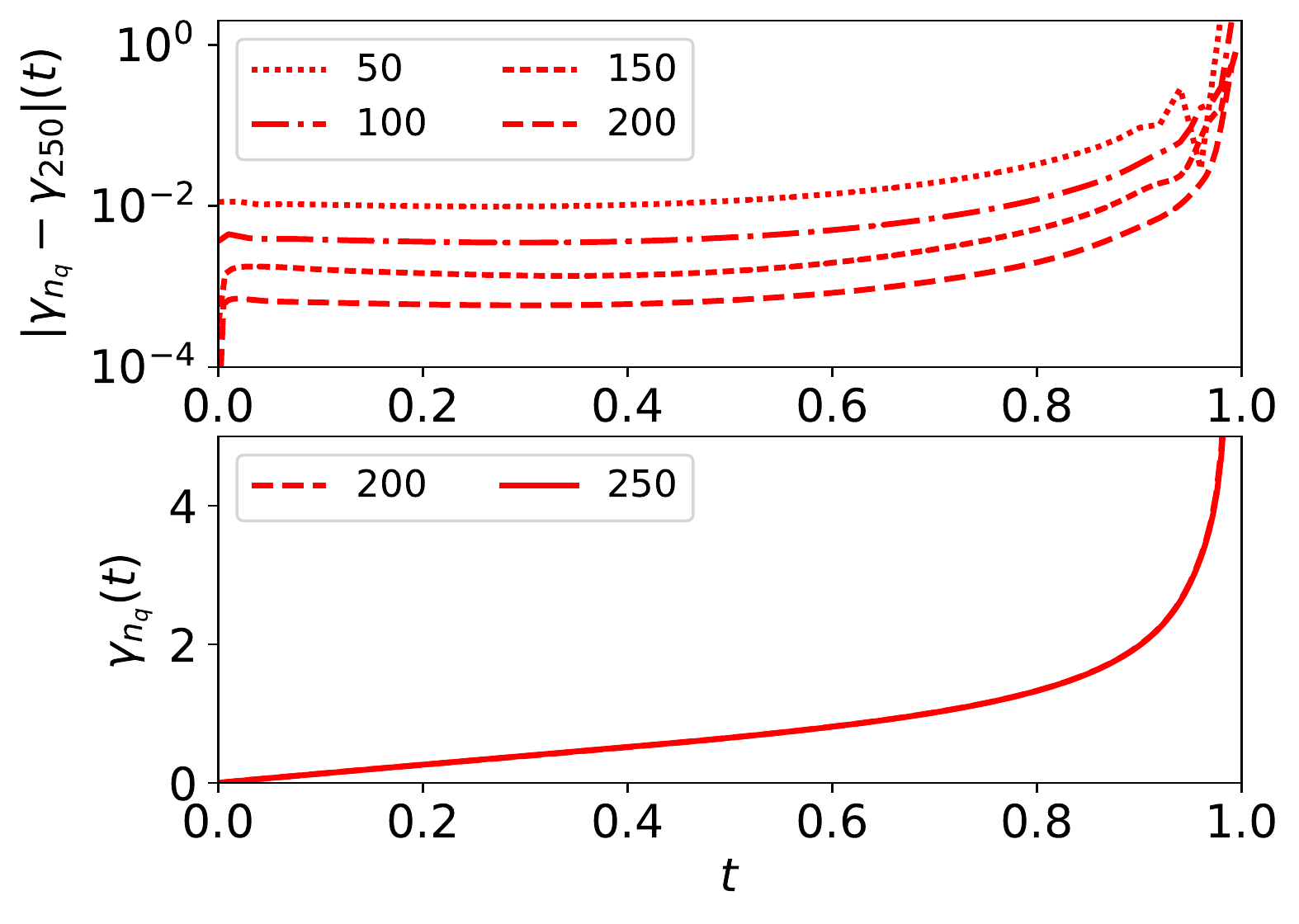}
\hspace{.1cm}
\includegraphics[width=.45\textwidth]{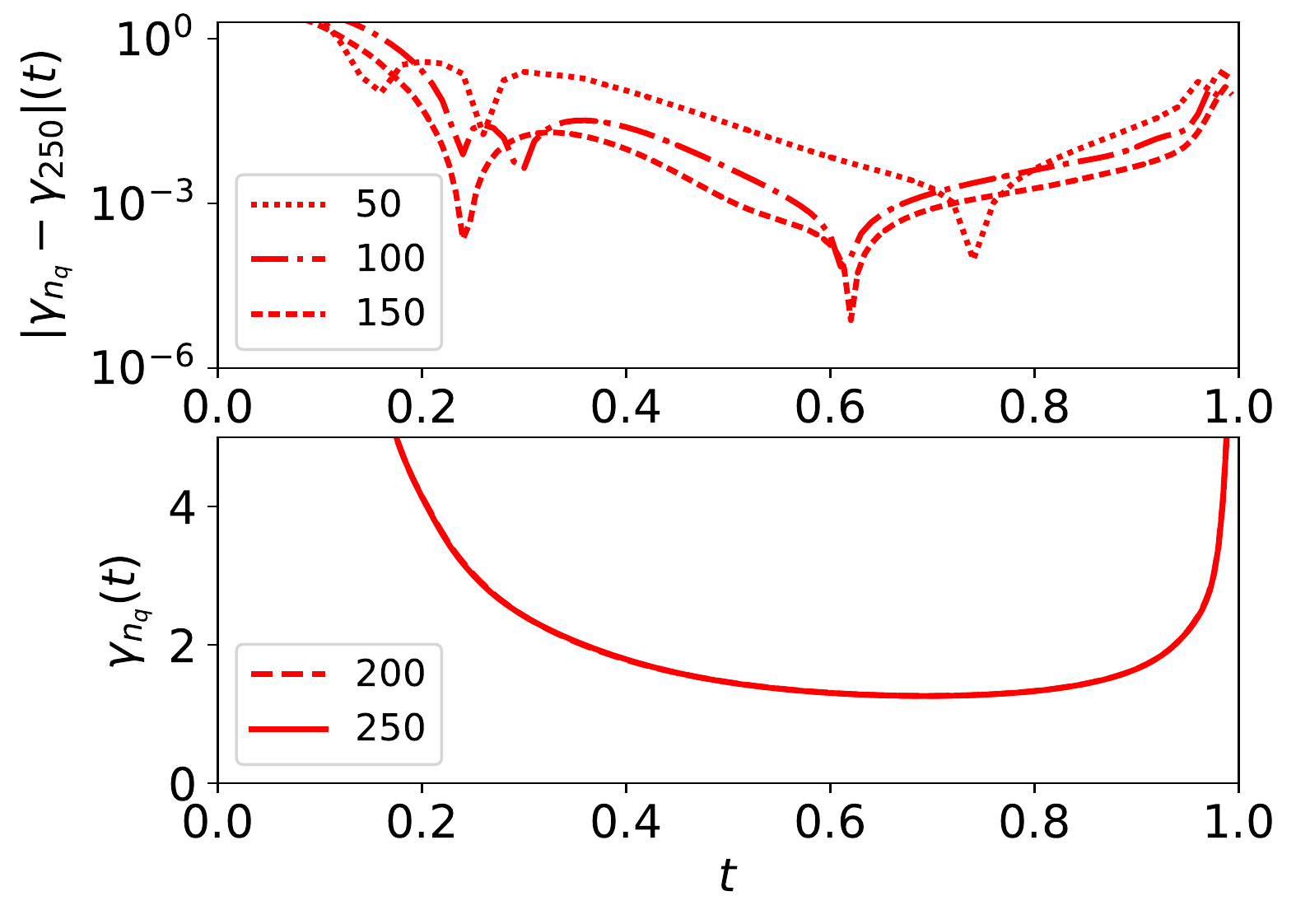}
\vspace{.2cm}
\caption{Numerical solutions  $\gamma_*$ of the extended variational
  principle \eqref{eq:VarPrinciple} or the 2-spin (SK) model (left) and 3-spin model (right). We
  discretize the function  $\gamma$ using $\nq = 50, 100, 150, 200, 250$ steps (in physics language, this corresponds to $\nq$-steps of RSB), and denote the resulting solution by $\gamma_{\nq}$. Top plots:
  difference between $\gamma_{\nq}$ and $\gamma_{250}$, indicating the order of magnitude of the discretization error.
  Lower plots: solutions for the finer discretization $\nq = 200, 250$.}
\label{fig:fop_ext}
\end{figure}

Figure~\ref{fig:fop_ext} (left frame) shows the optimizers of the variational problem
corresponding to $\he_{\alg}(\para)$ for the $2$-spin and $3$-spin models
for a few values of $\nq$.  For the $2$-spin model, we observe that the optimizer satisfies $r_1 < r_2 < \cdots  <
r_{\nq}$.
Since $\br\mapsto \Par_{\para}(\br)$ is a convex function, we conclude that $\he_{\alg}(\para)=\he_{\opt}(\para)$.
By solving problem \eqref{eq:finite} for $\nq\in\{50, 100,  150, 200\}$, we obtain the estimates: 
\begin{align}
\he_{\alg}^{\2spin}= \he_{\opt}^{\2spin}= 0.763168\pm 0.000002\, . \label{eq:2spinPred}
\end{align}
This compares well with the value published in \cite{crisanti2002analysis} ($0.76321 \pm 0.00003$) and the high precision extrapolation
of \cite{schmidt2008replica} ($0.763166726$).   

In the case of the 3-spin model, the situation is qualitatively different, see Figure~\ref{fig:fop_ext}, right frame.
We observe that the constraint $r_1 \le r_2 \le \cdots \le r_{\nq}$
in the definition of $\he_{\opt}(\para)$, see Eq.~\eqref{eq:finite}, is active:  the optimal vector of weights $\br^{\opt}$ is such that $r^{\opt}_1 = r^{\opt}_2 =\cdots =r^{\opt}_{n_0} < r^{\opt}_{n_0+1} < \cdots < r^{\opt}_{\nq}$ for some $1 <n_0 < \nq$.
On the other hand, the optimizer $\br^{\alg}$ in the definition of $\he_{\alg}(\para)$
is such that $r^{\alg}_1 > r^{\alg}_2> \cdots > r^{\alg}_{n'_0} < r^{\alg}_{n'_0+1} < \cdots < r^{\alg}_{\nq}$.
Consequently the two optimization problems in Eq.~\eqref{eq:finite} have different values.
  We also observe that the optimization problem \eqref{eq:finite} for $\he^{\3spin}_{\opt}(\para)$, while still convex, is
  harder than in the 2-spin case (convergence of projected gradient descent is slower). 
  In order to accelerate convergence, we use a discrete ansatz for  $\gamma$ that takes explicitly account of the overlap gap between $0$ and $q_0$:
  \begin{align}
    \gamma(t)  = r_{-1} \one_{[0,q_{0})}(t) + \sum_{i=0}^{\nq-1} r_i \one_{[q_{i},q_{i+1})}(t),
    \end{align}
with $q_i = q_0 + (1-q_0)i/\nq$ and $r_{-1} < r_0 < r_1 <\cdots < r_{\nq-1}$. 
We run projected gradient descent w.r.t.\ the variables $(r_{-1},r_0,\cdots,r_{\nq-1})$, with $q_0$ fixed.
If $r_0 = r_{-1}$ at some iteration of the algorithm, we make the update $q_0 \leftarrow q_0 + (1-q_0)/\nq$ and optimize over the vector $(r_1,\cdots,r_{\nq-1}) \in \reals^{\nq-1}$, and then set $\nq \leftarrow \nq-1$.
The algorithm is run until the value of $q_0$ stabilizes and the gradient of $\Par_{\para}(\br)$ is smaller than $3 \cdot 10^{-5}$. We numerically find
\begin{align}
  \he_{\alg}^{\3spin}= 0.8004\pm 0.0002\, , \qquad \qquad  \he_{\opt}^{\3spin}=  0.8132 \pm 0.0001\, .\label{eq:3spinPred}
  \end{align}
  %
%

\subsection{Calculation of an approximate optimizer}
\label{sec:Optimization}

We consider the version of the algorithm defined by Eqs.~\eqref{eqs:AlgUV_1} to \eqref{eqs:AlgUV_5},
where $u$, $v$ are given in Eq.~\eqref{eq:OptimalUV}, with a slight modification.
As mentioned in Section \ref{sec:Description}, the increments $(\de\bz^t)_{t\in\T_{\delta}}$ generated by the algorithm are
asymptotically orthogonal. We partially enforce this at finite $n$ through an explicit orthogonalization step.
Further, the analysis of the algorithm of Section \ref{sec:Description}, see \cite{alaoui2020optimization},
implies that $\|\de\bz^t\|_2^2= N(\xi'(t+\delta)-\xi'(t))+o(N)$ and $\|\bm^t\|_2^2 = Nt+o(N)$.
We also enforce these normalizations explicitly.

 Namely, we fix an integer $k \ge 1$ and, for each step $t$, we perform the following additional computations:
\begin{enumerate}
\item  We orthogonalize the increment $\de\bz^{t} := \bz^{t+\delta}-\bz^{t}$ with the previous $k$ increments $\de\bz^{t-\delta}$, $\de\bz^{t-2\delta}$, $\cdots,\de\bz^{t- k\delta}$, and normalize it to have norm $\sqrt{N(\xi'(t+\delta) - \xi'(t))}$:
 \begin{align}
\de \bz^{t} &\longleftarrow \proj_{\bH_{t,k}^{\perp}}\big(\de\bz^{t}\big),\\
  \de\bz^{t} &\longleftarrow  \sqrt{N(\xi'(t+\delta) - \xi'(t))}\, \frac{\de\bz^{t}}{ \|\de\bz^{t}\|_2}\, ,
 \end{align}
 where $\bH_{t,k} = \textup{span}\big(\de\bz^{t-\delta}, \de\bz^{t-2\delta}, \cdots,\de\bz^{t- k\delta}\big)$.
 \item We normalize the vector $\bm^{t}$, computed via  iteration \eqref{eqs:AlgUV_3}, to have norm $\sqrt{tN}$:
 \begin{equation}
  \bm^{t} \longleftarrow  \sqrt{N t}\, \frac{\bm^{t}}{\|\bm^{t}\|_2}\, .
 \end{equation}
\end{enumerate}         
Enforcing these constraints leads to better numerical stability of the algorithm.
In our experiments we use $k=5$.
We use the numerical solution of the variational principle \eqref{eq:VarPrinciple} described in the previous
section to compute the nonlinearities $u$ and $v$ via Eq.~\eqref{eq:OptimalUV}.
For the $2$-spin model, we round the algorithm by taking $\bsigma^{\alg}=\sign(\bm^{t=1})$. For the $3$-spin
model, we find it more effective to compute the energy $H_N(\bsigma^t)$, for $\bsigma^t= \sign(\bm^{t})$ at each iteration,
and return the vector $\bsigma^t$ that maximizes it.

We run this algorithms on random instances of the 2-spin and 3-spin models,  for various values of $N$ and the step size $\delta$ and record the achieved energy $e_{\bG}(N,\delta):=H_N(\bsigma^{\salg})/N$
(here $\bG$ refers to the randomness in the Hamiltonian $H_N$).
We use the values $N \in\{ 750,100,1250,2000,4000\}$ for the 2-spin model and $N \in\{ 400,500,600,750\}$ for the 3-spin model. For each value of $N$ we use $\delta \in \{ 4/N, 3/N, 2/N, 1/N\}$ for both models.

We are interested in the behavior as $N\to\infty$, $\delta\to 0$. In
order to extract these asymptotics, we evaluate the median
$\he(N,\delta)$ over $\ns=100$ independent realizations for the SK model, and $\ns=20$ realizations for the 3 spin model.
We then 
perform least squares linear  regression using the model
\begin{equation}\label{eq:regression}
\he_{N,\delta} = e_0 + \beta_1 N^{-a} + \beta_2 \delta\, .
\end{equation}
We choose the exponent $a$ on the basis of earlier statistical physics literature. For
the $2$-spin model  we use $a=2/3$, which is believed to capture the behavior
of the optimum at finite $N$, namely $e_{\opt}(N) = e_{\opt}+c_1\, N^{-2/3}+o(N^{-2/3})$
\cite{aspelmeier2008finite}. The same behavior is follows from the Tracy-Widom law if we replace
the constraint $\bsigma\in\{-1,+1\}^N$ with $\bsigma\in\S^{N-1}(\sqrt{N})$ \cite{Guionnet}. For the
$3$-spin model we use $a=1$, which is expected to be the correct behavior for the optimum in  models
with 1-RSB \cite{derrida1980random}. Although the 3-spin model at zero temperature is expected
to be FRSB, the structure of the order parameter is very close to 1-RSB\footnote{The optimal $\gamma$ in Eq.~\eqref{eq:original} is expected to be constant on an interval $[0,q_{0})$ and strictly increasing over $[q_0,1)$ with a jump discontinuity at $q_0$, with $q_0$ being rather close to 1: low precision numerics indicate $q_0 \in [0.79,0.81]$.}, and hence we expect $N^{-1}$ corrections to dominate, at least at moderate sizes.
We find 
\begin{align}\label{eq:reg_values}
\begin{split}
e_{0,\textup{2-spin}} = 0.7630738, \qquad & \qquad e_{0,\textup{3-spin}} = 0.7999287\, ,\\
\beta_{1,\textup{2-spin}} = -0.87532, \qquad & \qquad \beta_{1,\textup{3-spin}} = -4.65328\, ,\\
\beta_{2,\textup{2-spin}} = 0.59323, \qquad & \qquad \beta_{2,\textup{3-spin}} = -0.30750\,. 
\end{split}
\end{align}
The above numerics are in striking agreement with the theoretical predictions of Eqs.~\eqref{eq:2spinPred} and \eqref{eq:3spinPred}
which were obtained using the variational principle \eqref{eq:VarPrinciple}.

\begin{figure}
\centering
\includegraphics[width=.47\textwidth]{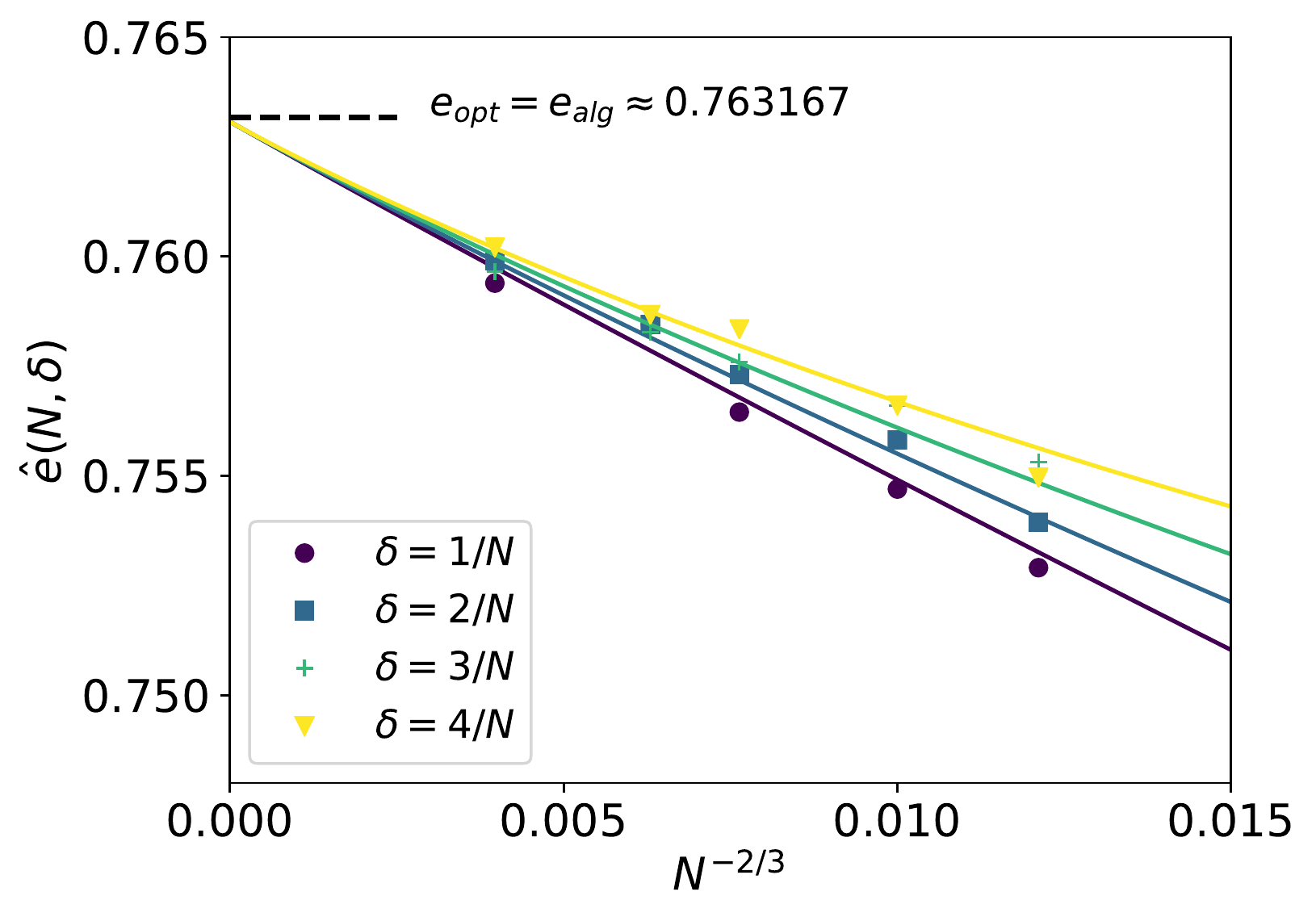}
\hspace{.1cm}
\includegraphics[width=.47\textwidth]{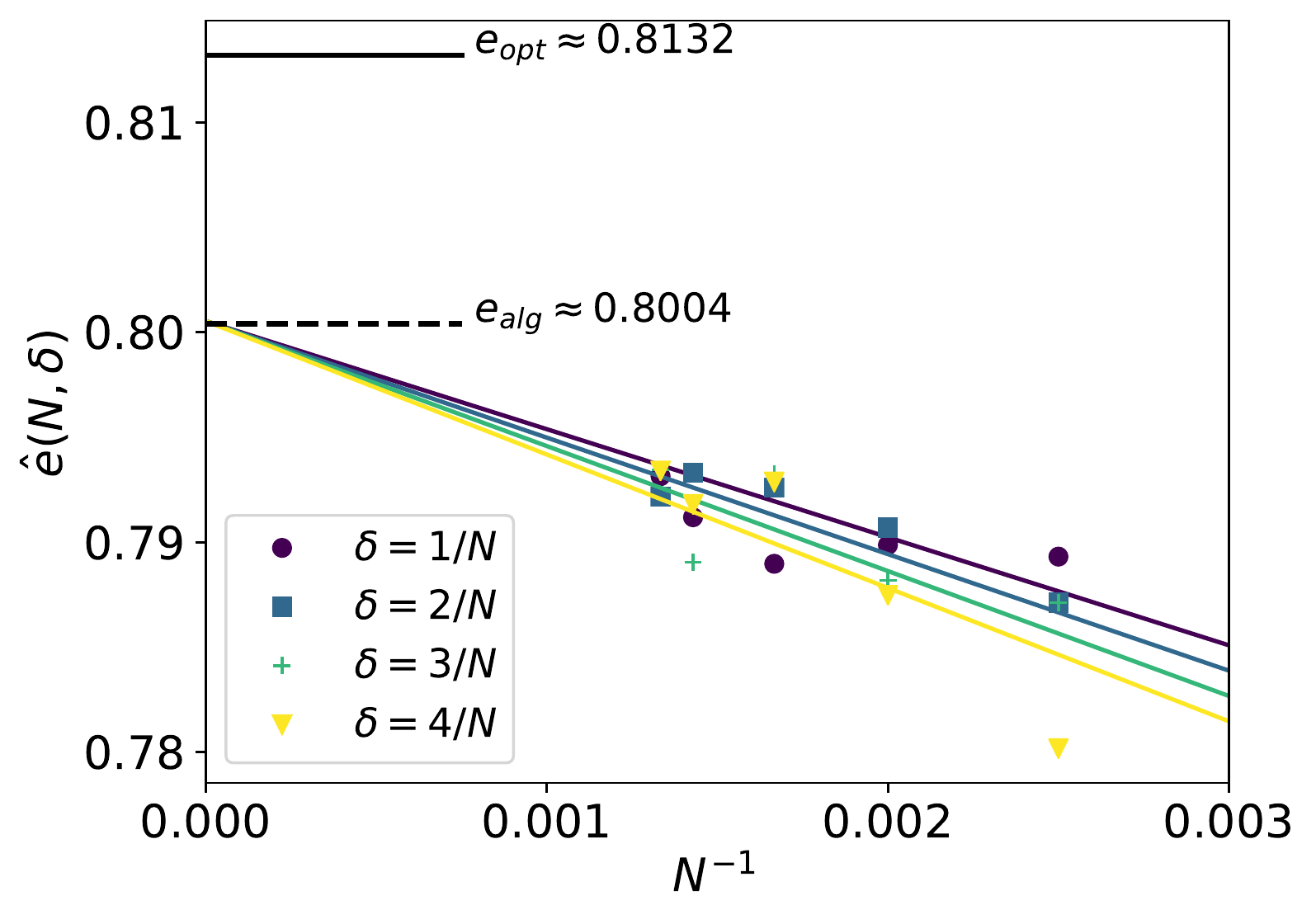}
\vspace{.2cm}
\caption{Energies achieved by the algorithm of Section \ref{sec:Optimization} for several values of the system size
  $N$ and the step size $\delta$. Left: SK model. Right: 3-spin model.
  Data points corresponds to medians over $\ns$ independent realizations ($\ns=100$ for SK, $\ns=20$ for $3$-spin model),
  with different symbols corresponding to different choices of $\delta$ (see legend). Lines corresponds to the result of the linear
  regression \eqref{eq:regression}, with each line corresponding to a different choice of $\delta$.}
\label{fig:reg_N_delta}
\end{figure}

In Figure~\ref{fig:reg_N_delta} we plot the empirical medians $\he_{N,\delta}$ as a function of $N^{-a}$, together
with the curves $e_0+\beta_1N^{-a}+\beta_2(k/N)$, with $k\in\{1,2,3,4\}$ corresponding to the regression \eqref{eq:regression}.
The model  \eqref{eq:regression} seems in agreement with the data, a finding that deserves further investigation.
Notice in particular that it is a priori unclear that the exponent $a=2/3$ that describes the convergence of the optimum energy to its
thermodynamic limit should also be appropriate for the behavior of the algorithm.

\subsection{Algorithmic threshold}

Our simulations confirm that the IAMP algorithm achieves the theshold energy
$e_{\alg}$ defined by the variational principle \eqref{eq:VarPrinciple}. Further, in models with overlap gap
the latter is a constant factor below the asymptotic value of the global optimum $e_{\opt}$.
A very interesting question is whether there exists a polynomial-time algorithm that can
achieve ---with high probability--- a better energy value than $e_{\alg}$.

While this question is widely open, physicists have devoted considerable attention
to the study of one specific algorithm: Glauber dynamics (a.k.a. Gibbs sampling).
This is a reversible Markov chain to sample from the Boltzmann distribution $\mu_{N,\beta}(\bsigma)$.
We regard it as an optimization algorithm by initializing it uniformly at random, and running it at a fixed,
large value of $\beta$ independent of $N$.
While no exact treatment of Glauber dynamics exists for Ising spin glasses, a conjecture was put forward in
\cite{rizzo2013replica} based on analysis of the energy landscape. Denoting by $e_{\Glauber}(\beta,t,N)$ the average
energy achieved by Glauber dynamics after $Nt$ updates on an instance of size $N$, we expect
$\lim_{\beta\to\infty}\lim_{t\to\infty}\lim_{N\to\infty}e_{\Glauber}(\beta,t,N) = \he_{\Glauber}$ (this limit is not expected to change if
$t$ scales polynomially in $N$).
The analysis of \cite{rizzo2013replica}  suggests\footnote{The value quoted here is an extrapolation
  to zero temperature from the non-zero temperature results of \cite{rizzo2013replica}. While 
  \cite{rizzo2013replica} warns against such extrapolation, the result
  is also consistent with the zero temperature 2RSB calculation of \cite{montanari2003nature},
  and the difference between $\he^{\3spin}_{\Glauber}$  and $\he^{\3spin}_{\alg}$ is large enough
  that it might overcome the extrapolation error.}:
\begin{align}
  \he^{\3spin}_{\Glauber}\lesssim 0.788 \, .
\end{align}
This is smaller than $\he^{\3spin}_{\alg}\approx 0.8004(2)$ (cf. Eq.~\eqref{eq:3spinPred}): the simple Glauber
dynamics algorithm  does not surpass the algorithmic threshold. On the other hand,
simulated annealing ---with a different temperature schedule--- is believed to achieve a better energy than
simple Glauber dynamics initialized uniformly at random \cite{montanari2004cooling,folena2020rethinking}:
it would be interesting to understand how does simulated annealing compare with the threshold $\he^{\3spin}_{\alg}$.

These questions can be studied more easily in the context of spherical models (recall that in this case the constraint
$\bsigma\in\{-1,+1\}^N$ is replaced by $\bsigma\in \S^{N-1}(\sqrt{N})$), and considering Langevin instead of Glauber
dynamics (in the zero-temperature limit, the former reduces to gradient flow).
We carry out this analysis in Appendix \ref{app:Spherical} for two specific examples of the generating function
$\xi(\,\cdot\,)$, both resulting in RSB with overlap gap. Our findings are consistent with the conclusions 
for the Ising case that we discussed above. Namely, the energy value $e_{\alg}$ achieved by the algorithms of
\cite{subag2018following,alaoui2020optimization} is superior to the threshold $e_{\th}$  of gradient flow.
In a special case, $\xi(x) = \frac{x^3}{2}+\frac{x^4}{2}$, we can also compare $e_{\alg}$
with the threshold $e_{\cool}$  of a different annealing algorithm studied in \cite{folena2020rethinking}.
We find, again, $e_{\alg}>e_{\cool}> e_{\th}$: the  algorithms of
\cite{subag2018following,alaoui2020optimization} is superior also to this modified approach.

\subsection{Pairs of near-optima}

\begin{figure}
\centering
\includegraphics[width=.65\textwidth]{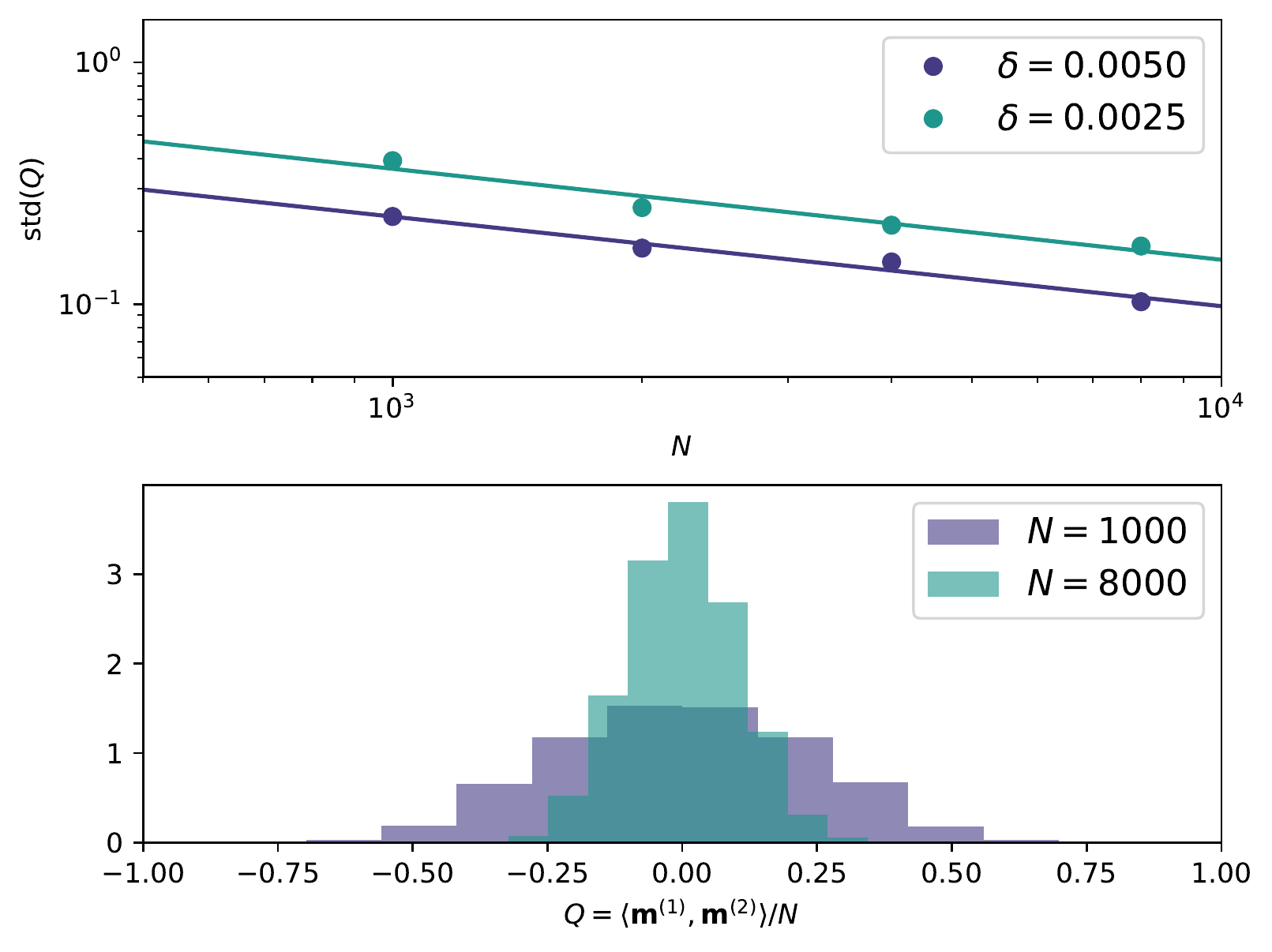}
\caption{Correlation between the near optima produced by two executions of the IAMP  algorithm on the same SK Hamiltonian,
  with independent  initializations. 
  Bottom plot: histograms of  $Q=\<\bm^{(1)},\bm^{(2)}\>/N$ for $\delta=1/200$ and two values of $N$.
  Top plot: standard deviation of $Q$ as a function of $N$. Lines are least squares fits sing 
  $\log{\rm std}(Q) =  c-\alpha \log N$. We obtain  $\alpha\approx
  0.37$. In both cases we average over $100$ independent initializations (hence $\binom{100}{2}$ pairs) and $2$ realizations of the Hamiltonian.}
\label{fig:bifurcation}
\end{figure}
One striking prediction of spin glass theory is that, for any
$q\in \cQ= \supp(\gamma_{\opt})$ and any $\eps>0$,
with high probability, there exist $\bsigma^{1}, \bsigma^{2} \in \cS_N(\eps)$
such that $\<\bsigma^{1},\bsigma^{2}\>/N\approx q$ \cite{auffinger2020sk}. (Here $\gamma_{\opt}$
is the optimizer of Parisi formula \eqref{eq:original}, and $\supp(\gamma_{\opt})$ is the closure
of the set of points at which $\gamma_{\opt}$ is strictly increasing.)

It is natural to wonder whether the same can be achieved by our IAMP algorithm.
In particular, in cases ---such as for the SK model--- in which we can find a near optimum, can we also find pairs
of near optima $\bsigma^{1},\bsigma^{2}$  at nearly all possible values of
$|\<\bsigma^{1},\bsigma^{2}\>|/N\in [0,1]$?

A possible way to obtain such a pair of near optima would be the following.
We run the algorithm up to a fixed (non-random) time $t_0\in \T_{\delta}$, then at time $t_0$ we bifurcate the trajectory into two parallel copies indexed by $a\in\{1,2\}$: $(\bz^{(a),t},\bm^{(a)a,t})$. The update for $\bm^{(a),t_0}$ is performed at random.
Namely, $\bz^{(1),t}=\bz^{(2),t}$ and $\bm^{(1),t}=\bm^{(2),t}$ for all $t \le t_0-\delta$. We draw $\bg^{(1)},\bg^{(2)}\sim_{iid}\normal(0,\id_N)$ and set
\begin{align}
  \bm^{(a),t}& =\begin{cases}
      \bm^{(a),t-\delta}+\sqrt{\delta} \bg^{(a)}&\;\;\;\;\;\;\mbox{if $t=t_0$,}\\
    \bm^{(a),t-\delta}+u_{\delta}(t-\delta;\bx^{(a),t-\delta})\odot (\bz^{(a),t}-\bz^{(a),t-\delta})&\;\;\;\;\;\;\mbox{if $t \ge t_0+\delta$.}
    \end{cases}
\end{align}
We expect this to produce a pair of near optima $\bm^{(1)}:=\bm^{(1),1}$, $\bm^{(2)}:=\bm^{(2),1}$
with $\<\bm^{(1)},\bm^{(2)}\>/N\approx t_0$ (taking the limit $\delta\to 0$ \emph{after} $N\to\infty$).

We carry out this test for $t_0=0$, i.e.\ running the algorithm from two independent initializations.
To make the effect more visible, we choose the initialization with larger variance:
$\bz^{\delta_0}\sim\normal(0,\delta_0\id_N)$, where $\delta_0= 10\,\delta$. 
The results of this experiment are summarized in Fig.~\ref{fig:bifurcation}.
For each of $N\in\{1000,2000,4000,8000\}$, $\delta\in\{1/200,1/400\}$, we generate $n=2$ realizations
of the SK Hamiltonian. For each realization, we run the algorithm from $M=100$ independent initializations.
We estimate the distribution of the scalar product between algorithm outputs $Q=\<\bm^{(1)},\bm^{(2)}\>/N$
(for the same Hamiltonian) by taking the empirical distribution over the $n\binom{M}{2}$ pairs corresponding to the same Hamiltonian. We observe that this distribution is unimodal and centered around $0$. The width of this distribution shrinks
with $N$ (for any fixed $\delta$) and data are consistent to the width vanishing as ${\rm std}(Q) \propto N^{-\alpha}$, $\alpha = 0.37$.

\section{Relation with physics}

As described in Section \ref{sec:Introduction}, the construction of
the algorithm is motivated by the picture of the
structure of ancestor states first developed in \cite{mezard1984nature,mezard1985microstructure}.
We therefore expect that the vector $\bm^t$ generated at iteration $t$ to have similar properties as the
magnetization vector $\bm^{\gamma}$ of a typical ancestor state with $\|\bm^{\gamma}\|^2_2/N\approx t$.
Here we investigate  two specific consequences of this picture: $(i)$ The generalized TAP free energy should be approximately
constant during the algorithm execution, and equal to the final energy achieved; $(ii)$ The generalized TAP equations should be
approximately satisfied at each iteration.

\begin{figure}
\centering
\includegraphics[width=.45\textwidth]{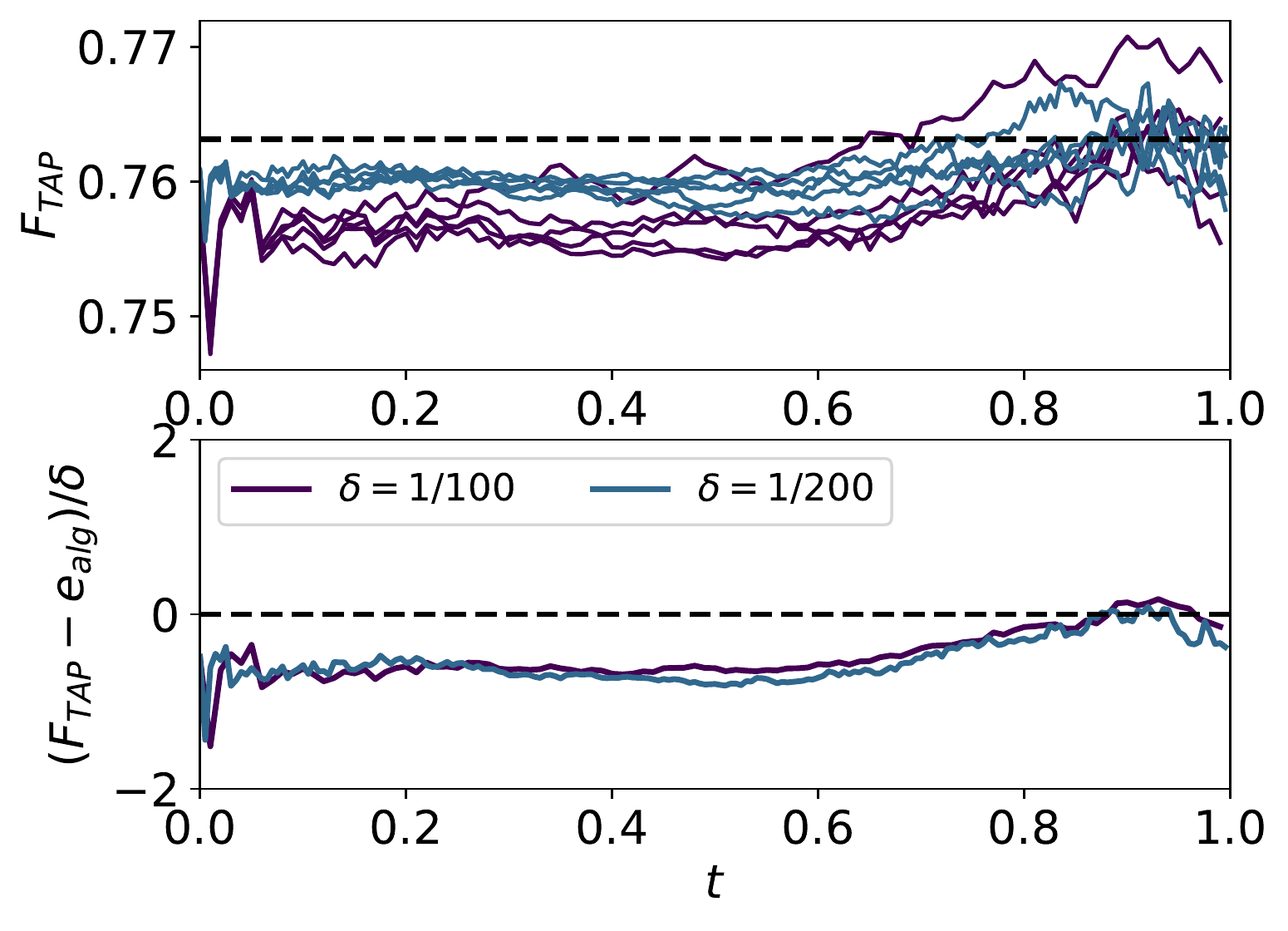}
\hspace{.1cm}
\includegraphics[width=.45\textwidth]{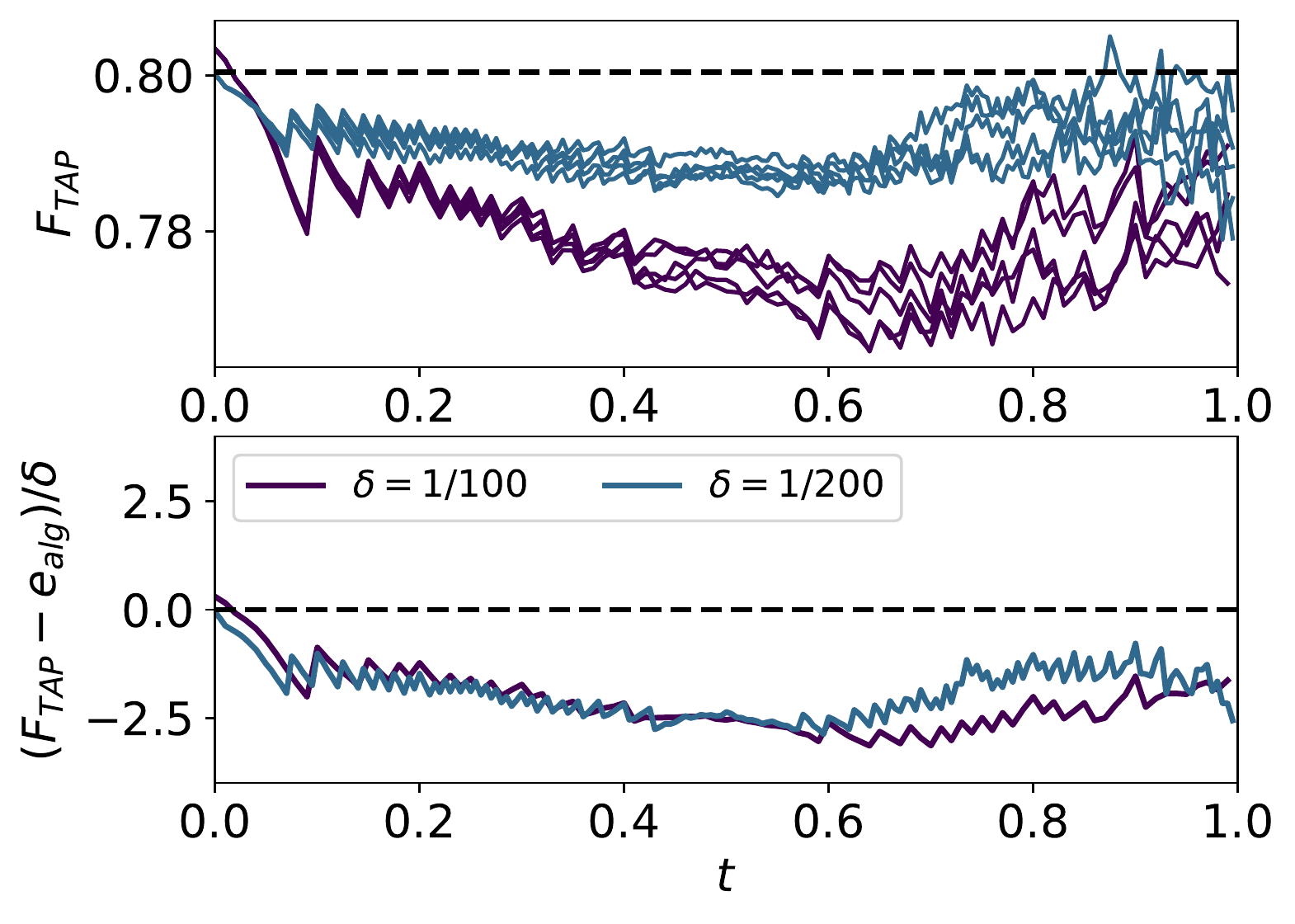}
\vspace{.2cm}
\caption{Evolution of the generalized TAP free energy during the execution of the algorithm, indexed by $t\in [0,1]$.
  Here $N=2000$ and we study the dependence on the step size $\delta$, for the $2$-spin model (left) and the $3$-spin
  model (right).  Top plots: trajectories of the free energy for $5$ independent realizations (colors correspond to the step size).
  Bottom plots, rescaled gap between the TAP free energy and the algorithmic threshold $e_{\alg}$. The collapse of curves in the
  bottom plots suggests that $\Delta_t(\delta,\infty)=\lim_{N\to\infty}(\cuF_{\sTAP}(\bm^t)/N -e_{\alg})=\Theta(\delta)$.}
\label{fig:tap}
\end{figure}

Each ancestor state $\gamma$ can be assigned a free energy which is the log partition function restricted to the set of configurations in that state. The generalized TAP free energy expresses this as a function of the magnetization vector $\bm^{\gamma}$.
The TAP free  energy associated to a magnetization vector $\bm \in \R^N$ with $\|\bm\|^2/N = t$ reads \cite{chen2018generalized,chen2019generalized}:
\begin{align}
  \cuF_{\sTAP}(\bm) &= H_N(\bm)+\sum_{i=1}^N\Lambda_{\gamma}(t,m_i)-\frac{N}{2}\int_t^1s\xi''(s)\gamma(s)\,\de s\, ,\\
  \Lambda_{\gamma}(t,m) &:= \inf_{s\in\reals}\big[\Phi_{\gamma}(t,x)-mx\big]\, .
\end{align}
In Figure \ref{fig:tap} we report the results of numerical experiments in which we evaluate $\cuF_{\sTAP}(\bm^t)/N$
along the iterates of the algorithm. These results are consistent with the hypothesis that
$\cuF_{\sTAP}(\bm^t)/N = e_{\alg}+ \Delta_t(\delta,N)$, where $\Delta_t(\delta,N)$ is an error that vanishes as $N\to\infty$,
and $\delta\to 0$. In particular the data in Fig.~\ref{fig:tap}  suggest $\Delta_t(\delta,\infty)  = \Theta(\delta)$. For completeness we included in Appendix~\ref{sec:TAP_proof} a proof that $\cuF_{\sTAP}(\bm^t)/N$ is indeed constant and equal to $e_{\alg}$ for all $t \in [0,1)$, in the limit $N\to\infty$ followed by $\delta \to 0$.

In our experiments we set $\gamma=\gamma_*$, the optimizer in the variational principle
\eqref{eq:VarPrinciple}. For the SK model, this coincides with the optimizer of Parisi formula
\eqref{eq:original}, and hence it is the right prescription for the dominant TAP states \cite{chen2018generalized,chen2019generalized}. On other hand, for the 3-spin model, the optimizer $\gamma_*$ is non-monotone. This case is not
covered by earlier theories in physics or mathematics.

\begin{figure}
\centering
\includegraphics[width=.65\textwidth]{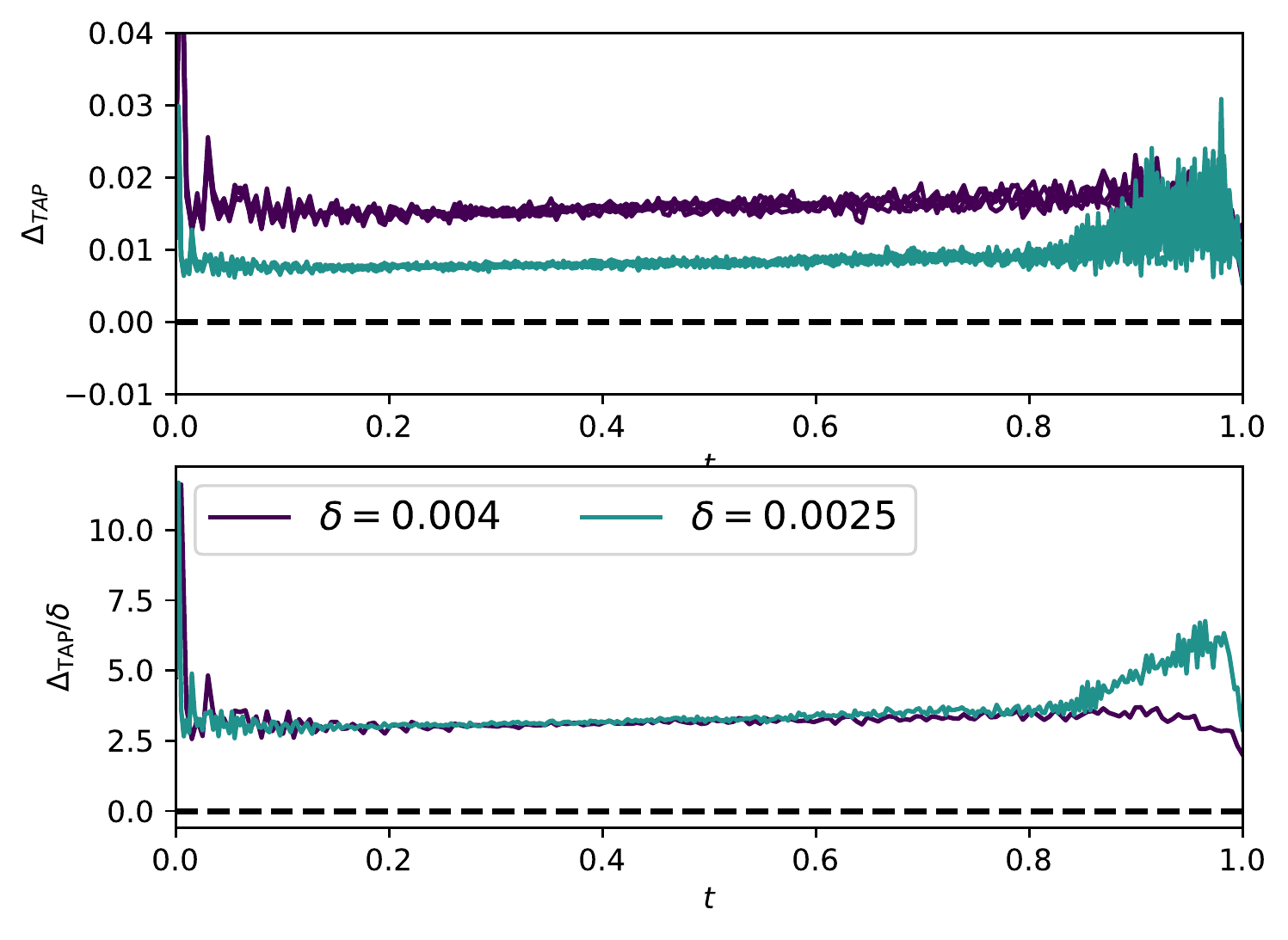}
\caption{Evolution of the generalized TAP equations during the execution of the algorithm, indexed by $t\in [0,1]$.
  Here $N=4000$ and we study the dependence on the step size $\delta$, for the SK ($2$-spin) model.
  Top plots: trajectories of the error $\tilde\Delta_{t}(\delta,N0$ for  $5$ independent realizations (colors correspond to the step size).
  Bottom plots, rescaled error $\tilde\Delta_{\sTAP}(\delta,N)/\delta$ averaged over the $5$ trajectories.}
\label{fig:tapeq}
\end{figure}

The magnetization vectors of ancestor states are approximate stationary points of the generalized TAP free energy.
The stationarity conditions are equivalent  to the following TAP equations (for $\|\bm\|_2^2/N=t$):
\begin{align}
   \bz&=\nabla H_N(\bm) -\bm\xi''(t)\int_t^1  \gamma(s) \,  \de s\, ,\label{eq:GenTAP1}\\
  \bm &= \partial_x\Phi_{\gamma}(t;\bz)\, . \label{eq:GenTAP2}
\end{align}
Writing these equations as $\bm = \bF(\bm;t)$, we plot in Fig.~\ref{fig:tapeq} the error
$\tilde\Delta_{t}(\delta,N):=\|\bm^t-\bF(\bm^t;t)\|_2^2/N$,
where $\bm^t$ is the vector produced by our algorithm. We observe that this value concentrates tightly about its average with respect to the realization. Further, the average decreases as $\delta\to 0$.
The rescaled plots suggest indeed $\tilde\Delta_{t}(\delta,\infty)= \Theta(\delta)$. 

Notice that the basic step of IAMP, cf. Eqs.\eqref{eq:ALG1},
\eqref{eq:ALG3} does not coincide with a simple iteration of the generalized TAP equations
\eqref{eq:GenTAP1}, \eqref{eq:GenTAP2}. This is different from the algorithm of \cite{bolthausen2014iterative},
that constructs
TAP solutions in the high temperature phase.


\section*{Acknowledgements}

This work was partially supported by the NSF grants CCF-1714305, CCF-2006489 and by the ONR grant
N00014-18-1-2729. 

\bibliographystyle{amsalpha}

\begin{thebibliography}{KMRT{\etalchar{+}}07}

\bibitem[ABM18]{addario2018algorithmic}
Louigi Addario-Berry and Pascal Maillard, \emph{The algorithmic hardness
  threshold for continuous random energy models}, arXiv:1810.05129 (2018).

\bibitem[ABMM08]{aspelmeier2008finite}
Timo Aspelmeier, Alain Billoire, Enzo Marinari, and Michael~A. Moore,
  \emph{{Finite-size corrections in the Sherrington--Kirkpatrick model}},
  Journal of Physics A: Mathematical and Theoretical \textbf{41} (2008),
  no.~32, 324008.

\bibitem[AC15]{auffinger2015parisi}
Antonio Auffinger and Wei-Kuo Chen, \emph{{The Parisi formula has a unique
  minimizer}}, Communications in Mathematical Physics \textbf{335} (2015),
  no.~3, 1429--1444.

\bibitem[AC17]{auffinger2017parisi}
\bysame, \emph{Parisi formula for the ground state energy in the mixed $p$-spin
  model}, The Annals of Probability \textbf{45} (2017), no.~6b, 4617--4631.

\bibitem[ACZ20]{auffinger2020sk}
Antonio Auffinger, Wei-Kuo Chen, and Qiang Zeng, \emph{{The SK Model Is
  Infinite Step Replica Symmetry Breaking at Zero Temperature}}, Communications
  on Pure and Applied Mathematics \textbf{73} (2020), no.~5, 921--943.

\bibitem[AGZ09]{Guionnet}
Greg~W. Anderson, Alice Guionnet, and Ofer Zeitouni, \emph{{An introduction to
  random matrices}}, Cambridge University Press, 2009.

\bibitem[AMS20]{alaoui2020optimization}
Ahmed~El Alaoui, Andrea Montanari, and Mark Sellke, \emph{Optimization of
  mean-field spin glasses}, arXiv:2001.00904 (2020).

\bibitem[BM11]{BM-MPCS-2011}
Mohsen Bayati and Andrea Montanari, \emph{{The dynamics of message passing on
  dense graphs, with applications to compressed sensing}}, IEEE Trans. on
  Inform. Theory \textbf{57} (2011), 764--785.

\bibitem[BMZ05]{braunstein2005survey}
Alfredo Braunstein, Marc M{\'e}zard, and Riccardo Zecchina, \emph{Survey
  propagation: An algorithm for satisfiability}, Random Structures \&
  Algorithms \textbf{27} (2005), no.~2, 201--226.

\bibitem[Bol14]{bolthausen2014iterative}
Erwin Bolthausen, \emph{{An iterative construction of solutions of the TAP
  equations for the Sherrington--Kirkpatrick model}}, Communications in
  Mathematical Physics \textbf{325} (2014), no.~1, 333--366.

\bibitem[CGPR19]{chen2019suboptimality}
Wei-Kuo Chen, David Gamarnik, Dmitry Panchenko, and Mustazee Rahman,
  \emph{Suboptimality of local algorithms for a class of max-cut problems}, The
  Annals of Probability \textbf{47} (2019), no.~3, 1587--1618.

\bibitem[CL04]{crisanti2004spherical}
Andrea Crisanti and Luca Leuzzi, \emph{Spherical 2+ p spin-glass model: An
  exactly solvable model for glass to spin-glass transition}, Physical review
  letters \textbf{93} (2004), no.~21, 217203.

\bibitem[CPS18]{chen2018generalized}
Wei-Kuo Chen, Dmitry Panchenko, and Eliran Subag, \emph{{The generalized TAP
  free energy}}, arXiv:1812.05066 (2018).

\bibitem[CPS19]{chen2019generalized}
\bysame, \emph{{The generalized TAP free energy II}}, arXiv preprint
  arXiv:1903.01030 (2019).

\bibitem[CR02]{crisanti2002analysis}
Andrea Crisanti and Tommaso Rizzo, \emph{{Analysis of the $\infty$-replica
  symmetry breaking solution of the Sherrington-Kirkpatrick model}}, Physical
  Review E \textbf{65} (2002), no.~4, 046137.

\bibitem[CS92]{crisanti1992sphericalp}
Andrea Crisanti and H-J Sommers, \emph{The sphericalp-spin interaction spin
  glass model: the statics}, Zeitschrift f{\"u}r Physik B Condensed Matter
  \textbf{87} (1992), no.~3, 341--354.

\bibitem[CS17]{chen2017parisi}
Wei-Kuo Chen and Arnab Sen, \emph{Parisi formula, disorder chaos and
  fluctuation for the ground state energy in the spherical mixed p-spin
  models}, Communications in Mathematical Physics \textbf{350} (2017), no.~1,
  129--173.

\bibitem[Der80]{derrida1980random}
Bernard Derrida, \emph{Random-energy model: Limit of a family of disordered
  models}, Physical Review Letters \textbf{45} (1980), no.~2, 79.

\bibitem[Der85]{derrida1985generalization}
\bysame, \emph{A generalization of the random energy model which includes
  correlations between energies}, Journal de Physique Lettres \textbf{46}
  (1985), no.~9, 401--407.

\bibitem[DMM09]{DMM09}
David~L. Donoho, Arian Maleki, and Andrea Montanari, \emph{{Message Passing
  Algorithms for Compressed Sensing}}, Proceedings of the National Academy of
  Sciences \textbf{106} (2009), 18914--18919.

\bibitem[EVdB01]{engel2001statistical}
Andreas Engel and Christian Van~den Broeck, \emph{Statistical mechanics of
  learning}, Cambridge University Press, 2001.

\bibitem[FFRT20a]{folena2020gradient}
Giampaolo Folena, Silvio Franz, and Federico Ricci-Tersenghi, \emph{Gradient
  descent dynamics in the mixed $ p $-spin spherical model: finite size
  simulation and comparison with mean-field integration}, arXiv preprint
  arXiv:2007.07776 (2020).

\bibitem[FFRT20b]{folena2020rethinking}
\bysame, \emph{Rethinking mean-field glassy dynamics and its relation with the
  energy landscape: The surprising case of the spherical mixed p-spin model},
  Physical Review X \textbf{10} (2020), no.~3, 031045.

\bibitem[GS14]{gamarnik2014limits}
David Gamarnik and Madhu Sudan, \emph{Limits of local algorithms over sparse
  random graphs}, Proceedings of the 5th conference on Innovations in
  theoretical computer science, ACM, 2014, pp.~369--376.

\bibitem[GS17]{gamarnik2017performance}
\bysame, \emph{{Performance of sequential local algorithms for the random
  NAE-K-SAT problem}}, SIAM Journal on Computing \textbf{46} (2017), no.~2,
  590--619.

\bibitem[JT16]{jagannath2016dynamic}
Aukosh Jagannath and Ian Tobasco, \emph{A dynamic programming approach to the
  parisi functional}, Proceedings of the American Mathematical Society
  \textbf{144} (2016), no.~7, 3135--3150.

\bibitem[KGV83]{kirkpatrick1983optimization}
Scott Kirkpatrick, C~Daniel Gelatt, and Mario~P Vecchi, \emph{Optimization by
  simulated annealing}, science \textbf{220} (1983), no.~4598, 671--680.

\bibitem[KMRT{\etalchar{+}}07]{krzakala2007gibbs}
Florent Krzakala, Andrea Montanari, Federico Ricci-Tersenghi, Guilhem
  Semerjian, and Lenka Zdeborov{\'a}, \emph{Gibbs states and the set of
  solutions of random constraint satisfaction problems}, Proceedings of the
  National Academy of Sciences \textbf{104} (2007), no.~25, 10318--10323.

\bibitem[KT87]{kirkpatrick1987p}
Theodore~R Kirkpatrick and Devarajan Thirumalai, \emph{p-spin-interaction
  spin-glass models: Connections with the structural glass problem}, Physical
  Review B \textbf{36} (1987), no.~10, 5388.

\bibitem[MM09]{MezardMontanari}
Marc M{\'e}zard and Andrea Montanari, \emph{{Information, Physics and
  Computation}}, Oxford, 2009.

\bibitem[Mon95]{monasson1995structural}
R{\'e}mi Monasson, \emph{Structural glass transition and the entropy of the
  metastable states}, Physical review letters \textbf{75} (1995), no.~15, 2847.

\bibitem[Mon19]{montanari2019optimization}
Andrea Montanari, \emph{{Optimization of the Sherrington-Kirkpatrick
  Hamiltonian}}, IEEE Symposium on the Foundations of Computer Science, FOCS,
  November 2019.

\bibitem[MP85]{mezard1985replicas}
Marc M{\'e}zard and Giorgio Parisi, \emph{Replicas and optimization}, Journal
  de Physique Lettres \textbf{46} (1985), no.~17, 771--778.

\bibitem[MP86]{mezard1986replica}
\bysame, \emph{A replica analysis of the travelling salesman problem}, Journal
  de Physique \textbf{47} (1986), no.~8, 1285--1296.

\bibitem[MPS{\etalchar{+}}84]{mezard1984nature}
Marc M{\'e}zard, Giorgio Parisi, Nicolas Sourlas, G~Toulouse, and Miguel
  Virasoro, \emph{Nature of the spin-glass phase}, Physical review letters
  \textbf{52} (1984), no.~13, 1156.

\bibitem[MPV87]{SpinGlass}
Marc M\'ezard, Giorgio Parisi, and Miguel~A. Virasoro, \emph{Spin glass theory
  and beyond}, World Scientific, 1987.

\bibitem[MPWZ02]{mulet2002coloring}
Roberto Mulet, Andrea Pagnani, Martin Weigt, and Riccardo Zecchina,
  \emph{Coloring random graphs}, Physical review letters \textbf{89} (2002),
  no.~26, 268701.

\bibitem[MPZ02]{mezard2002analytic}
Marc M{\'e}zard, Giorgio Parisi, and Riccardo Zecchina, \emph{Analytic and
  algorithmic solution of random satisfiability problems}, Science \textbf{297}
  (2002), no.~5582, 812--815.

\bibitem[MRT03]{montanari2003nature}
Andrea Montanari and Federico Ricci-Tersenghi, \emph{On the nature of the
  low-temperature phase in discontinuous mean-field spin glasses}, The European
  Physical Journal B-Condensed Matter and Complex Systems \textbf{33} (2003),
  no.~3, 339--346.

\bibitem[MRT04]{montanari2004cooling}
\bysame, \emph{Cooling-schedule dependence of the dynamics of mean-field
  glasses}, Physical Review B \textbf{70} (2004), no.~13, 134406.

\bibitem[MV85]{mezard1985microstructure}
Marc M{\'e}zard and Miguel~Angel Virasoro, \emph{The microstructure of
  ultrametricity}, Journal de Physique \textbf{46} (1985), no.~8, 1293--1307.

\bibitem[MZK{\etalchar{+}}99]{monasson1999determining}
R{\'e}mi Monasson, Riccardo Zecchina, Scott Kirkpatrick, Bart Selman, and
  Lidror Troyansky, \emph{Determining computational complexity from
  characteristic ‘phase transitions’}, Nature \textbf{400} (1999),
  no.~6740, 133.

\bibitem[Nis01]{NishimoriBook}
Hidetoshi Nishimori, \emph{{Statistical Physics of Spin Glasses and Information
  Processing: An Introduction}}, Oxford University Press, 2001.

\bibitem[Pan13a]{panchenko2013parisi}
Dmitry Panchenko, \emph{{The Parisi ultrametricity conjecture}}, Annals of
  Mathematics (2013), 383--393.

\bibitem[Pan13b]{panchenko2013sherrington}
\bysame, \emph{{The Sherrington-Kirkpatrick model}}, Springer Science \&
  Business Media, 2013.

\bibitem[PB14]{parikh2014proximal}
Neal Parikh and Stephen Boyd, \emph{Proximal algorithms}, Foundations and
  Trends in optimization \textbf{1} (2014), no.~3, 127--239.

\bibitem[Riz13]{rizzo2013replica}
Tommaso Rizzo, \emph{Replica-symmetry-breaking transitions and off-equilibrium
  dynamics}, Physical Review E \textbf{88} (2013), no.~3, 032135.

\bibitem[Sch08]{schmidt2008replica}
Manuel~J Schmidt, \emph{Replica symmetry breaking at low temperatures}, Ph.F.
  Thesis, 2008.

\bibitem[SK75]{sherrington1975solvable}
David Sherrington and Scott Kirkpatrick, \emph{Solvable model of a spin-glass},
  Physical review letters \textbf{35} (1975), no.~26, 1792.

\bibitem[Sub18]{subag2018following}
Eliran Subag, \emph{{Following the ground-states of full-RSB spherical spin
  glasses}}, {\sf arXiv:1812.04588} (2018).

\bibitem[Tal06]{talagrand2006parisi}
Michel Talagrand, \emph{{The Parisi formula}}, Annals of Mathematics (2006),
  221--263.

\bibitem[Tal10]{TalagrandBook}
Michel Talagrand, \emph{Mean field models for spin glasses}, Springer-Verlag,
  Berlin, 2010.

\end{thebibliography}

\newcommand{\etalchar}[1]{$^{#1}$}
\providecommand{\bysame}{\leavevmode\hbox to3em{\hrulefill}\thinspace}
\providecommand{\MR}{\relax\ifhmode\unskip\space\fi MR }
\providecommand{\MRhref}[2]{%
  \href{http://www.ams.org/mathscinet-getitem?mr=#1}{#2}
}
\providecommand{\href}[2]{#2}

\appendix

\newpage

\section{Further details on the solution of the variational principle}
\label{sec:NumDetails}

As mentioned in the main text, we solve the variational principle \eqref{eq:VarPrinciple}
numerically using the projected gradient method.
We give here the explicit expression of the gradient of $\Phi_{\gamma}(0,0)$ with respect to $\br$.

Recall from Section~\ref{sec:variational_principle} that using the Cole-Hopf transform with $\gamma(t) = \sum_{i=1}^{\nq} r_i \one_{t \in [q_i,q_{i+1})}$  we have 
\begin{equation*}
\Phi_{\gamma}(q_{i},x) = \frac{1}{r_i} \log \E \Big[\exp\Big(r_i \Phi_{\gamma}\big(q_{i+1},x + \sqrt{\xi'(q_{i+1}) - \xi'(q_i)}Z\big)\Big) \Big]~~~\forall 1\le i \le \nq,
\end{equation*}        
where $Z \sim \normal(0,1)$. Let $z_i(x) = x + \sqrt{\xi'(q_{i+1}) - \xi'(q_i)}Z$ for convenience.
The gradient with respect to the parameters $\br = (r_1,\cdots,r_{\nq})$ can be computed recursively as follows
\begin{align*}
  \frac{\partial\phantom{r_j}}{\partial r_{j}} \Phi_{\gamma}(q_{i},x) &= 
0 & ~\mbox{if}~ j \le i-1, \\
\frac{\partial\phantom{r_j}}{\partial r_{j}} \Phi_{\gamma}(q_{i},x) &= \frac{\E \Big[\Phi_{\gamma}(q_{i+1},z_i(x))e^{r_i \Phi_{\gamma}(q_{i+1},z_i(x))} \Big]}{\E \Big[e^{r_i \Phi_{\gamma}(q_{i+1},z_i(x))} \Big]} - \frac{1}{r_i^2} \log \E \Big[e^{r_i \Phi_{\gamma}(q_{i+1},z_i(x))} \Big] &~\mbox{if}~ j = i,\\
\frac{\partial\phantom{r_j}}{\partial r_{j}} \Phi_{\gamma}(q_{i},x) &= \frac{\E \Big[\frac{\de}{\de r_j}\Phi_{\gamma}(q_{i+1},z_i(x))e^{r_i \Phi_{\gamma}(q_{i+1},z_i(x))} \Big]}{\E \Big[e^{r_i \Phi_{\gamma}(q_{i+1},z_i(x))} \Big]} &~\mbox{if}~ j \ge  i+1. 
\end{align*}



\section{The case of the spherical model}
\label{app:Spherical}

Recall that the spherical model is defined by the Hamiltonian $H_N:\S^{N-1}(\sqrt{N})\to \reals$ being
a centered Gaussian process on the $N$-dimensional sphere with radius $\sqrt{N}$ ---denoted by $\S^{N-1}(\sqrt{N})$---
with covariance $\E\{H_N(\bsigma_1)H_N(\bsigma_2)\} = N\xi(\<\bsigma_1,\bsigma_2\>/N)$.
The spherical symmetry simplifies the treatment.

In this appendix we compare the behavior the algorithms of \cite{subag2018following,alaoui2020optimization},
with properties of the energy landscape, as derived within statistical physics. 
Our discussion will be mainly heuristic.

\subsection{Algorithm}

We can apply the algorithm of Eqs.~\eqref{eqs:AlgUV_1} to \eqref{eqs:AlgUV_5}. The optimal choice of
functions $u,v:[0,1]\to\reals\to\reals$
is given by the following optimization problem:
\begin{align}
  \mbox{maximize}&\;\; \cuE(u,v) := \int_0^1 \xi''(t) \E\big[u(t,X_t)\big] \de t\, ,\label{eq:DesignSpherical}\\
  \mbox{subj. to}&\;\; \xi''(t)\E\big[u(t,X_t)^2\big] = 1~~\mbox{for all}~t \in [0,1)\, , \nonumber
\end{align}
where it is understood that $X_t$ solves the SDE $\de X_t = v(t,X_t)\,\de t+\sqrt{\xi''(t)}\,\de B_t$. 
Unlike in the Ising case, we do not have any constraint on the terminal value $M_1$ of the martingale $M_t = \int_0^t \sqrt{\xi''(s)} u(s,X_s) \de B_s$. This is due to the fact that the Ising constraint 
$\bsigma\in\{-1,+1\}^N$ is replaced by an $\ell_2$ constraint $\|\bsigma\|_2^2 = N$,  which correspond to the
condition $\E\big[M_1^2\big]=1$. This in turn is a consequence of the constraint in the optimization problem~\eqref{eq:DesignSpherical}. The value
of the optimization problem \eqref{eq:DesignSpherical} corresponds to the value achieved by the algorithm.

Solving this optimization problem is straightforward. By Cauchy-Schwarz, the only optimizer is given by
$u(t,X_t) = 1/\sqrt{\xi''(t)}$. With these choices, the algorithm of Eqs.~\eqref{eqs:AlgUV_1} to \eqref{eqs:AlgUV_5} reduces to
\begin{align}
  \bz^{t+\delta} &= \nabla H_N(\bm^t)-\sum_{s\in \T_\delta\cap [\delta,t]}\sd_{t,s} \bm^{s-\delta}\, ,\label{eqs:AlgUV_1_s}\\
  \sd_{t,s}&:= -\xi''(\<\bm^t,\bm^{s-\delta}\>_N\big) \Big(\xi''(s)^{-1/2} \one_{s<t}-\xi''(s-\delta)^{-1/2}\Big)\, , \label{eqs:AlgUV_2_s}\\
  \bm^t& = \sum_{s\in \T_{\delta}\cap[0,t-\delta]}\xi''(s)^{-1/2} (\bz^{s+\delta}-\bz^s)\, .
\end{align}
The resulting value achieved is
\begin{align}
  \lim_{N\to\infty}\frac{1}{N}H_N(\bsigma^{\salg}) = e_{\alg} = \int_{0}^1 \sqrt{\xi''(t)}\, \de t\, .\label{eq:EalgSpherical}
\end{align}
This coincides with the value achieved by the algorithm of \cite{subag2018following}.


\subsection{Energy landscape}

The (zero temperature) Parisi functional can be written 
as an explicit function of the pair $\gamma,L$ where $\gamma \in\cuL$, and $L\ge \int_0^1\gamma(t) \de t$
\cite{crisanti1992sphericalp,crisanti2004spherical,chen2017parisi}:
\begin{align}
\Par(\gamma,L) & = \frac{1}{2}\int_0^1\left(\xi''(t)\Gamma(t) +\frac{1}{\Gamma(t)}\right)\de t\,,\label{eq:ParisiSpherical}\\
\Gamma(t) & := L-\int_0^t \gamma(s)\de s\,  .
\end{align}
Equivalently, we can view $\Par$ as a function of $\Gamma:[0,1]\to\reals_{\ge 0}$ which is continuous and non-increasing.
It is concave if and only if $\gamma$ is non-decreasing. 

As before, we will consider two variational principles associated with $\Par(\gamma,L)$:
\begin{align}
  e_{\alg}&:= \inf \big\{\Par(\gamma,L) :\; \gamma\in\cuL\, ,  \;\; L\ge \int_0^1\gamma(t) \de t\big\}\, ,
  \label{eq:SphericalVariationalAlg}\\
  e_{\opt}&:= \inf \big\{\Par(\gamma,L) :\; \gamma\in\cuL\, , \gamma \,\mbox{non-decreasing}\, \;\; L\ge \int_0^1\gamma(t) \de t\big\}\, .
  \label{eq:SphericalVariational}
\end{align}
It is easy to see that the first one is the dual of the maximization problem \eqref{eq:DesignSpherical}.
The minimum is achieved at $\Gamma(t) = 1/\sqrt{\xi''(t)}$, matching the value of  $e_{\alg}$
as per Eq.~\eqref{eq:EalgSpherical}. The second variational principle ---Eq.~\eqref{eq:SphericalVariational}--- yields the optimum value 
\begin{align}
  \lim_{N\to\infty}\frac{1}{N}\max_{\bsigma\in \S^{N-1}(\sqrt{N})}H_N(\bsigma) = e_{\opt} \, .\label{eq:EOptSpherical}
\end{align}

We will consider two specific  examples, i.e. two choices of the function $\xi(\,\cdot\,)$,
which we believe are representative of classes of models that differ in the structure of the solution
$(\gamma,L)$ of the variational problem \eqref{eq:SphericalVariational}:
\begin{enumerate}
\item \emph{One-step replica symmetry breaking (1RSB)}. In this case $\gamma(t)=\mu$ for $t\in [0,1)$ and $L>\mu$.
  Equivalently, $\Gamma(t) = L-\mu t$. As a prototype of this class, we study the `$3+4$' model:
  \begin{align}
    \xi(x) = \frac{1}{2}x^3+\frac{1}{2}x^4\, .
  \end{align}
  This pattern of replica symmetry breaking is the simplest possible, and has been studied in detail recently
  \cite{folena2020rethinking,folena2020gradient}.
  Hence, this model is particularly useful for comparing the value achieved by the algorithms studied here, and the features of the energy landscape. We will denote by $\mu_{\opt}$ the value of $\mu$ that corresponds to the minimizer of the variational
  principle \eqref{eq:SphericalVariational}.

  Geometrically, this structure of $\gamma$ corresponds to a one-level tree, whose leaves are well-separated pure states
  with $\|\bm^{\alpha}\|_1^2/N\approx 1$, and $\<\bm^{\alpha}\|_1^2/N\approx 1$ (the overlap distribution has support on
  $\{0,1\}$).
\item \emph{Full replica symmetry breaking with a single gap at $0$ (FRSB-1G).} In this case there exists $t_0\in (0,1)$
  such that $\gamma(t) = \mu$ for $t\in [0,t_0]$
  and $\gamma(t)$ strictly increasing and continuous for $t\in [t_0,1]$. Equivalently, $\Gamma(t) = a-\mu t$
  for $t\in [0,t_0]$, and $\Gamma(t)$ is strictly concave on $[t_0,1]$. As an example of this structure we use
  \begin{align}
    \xi(x) = \frac{1}{20}x^2+\frac{1}{6}x^3+\frac{1}{992} x^{32}\, .
  \end{align}
  This structure of RSB is interesting because it is believed to be quite generic, and in particular it is
  expected to be the same occurring for the pure $p$-spin Ising spin glass (i.e., the Ising spin glass with $\xi(x) = x^p$, $p\ge 3$).
  As before, we denote by $\mu_{\opt}$ the value of $\mu$ at the minimizer of the variational principle
  \eqref{eq:SphericalVariational}.
  
  From a geometric point of view, this corresponds to the ancestor states being well separated subsets
  of the sphere $\S^{N-1}(\sqrt{N})$. When restricted to such a set, the Gibbs measure presents full-replica
  symmetry breaking with no gaps. 
\end{enumerate}

In order to explore the energy landscape, we use the statistical physics method of `clones' first introduced in
\cite{monasson1995structural}.
This approach consists in finding a constrained optimum of the  functional $\Par(\gamma,L)$
\begin{align}
  \mbox{minimize}&\;\;\; \Par(\gamma,L)\, ,\nonumber\\
  \mbox{subject to}&\;\;\; \gamma\in \cuL\, ,\;\; \gamma \mbox{ non decreasing,}\label{eq:ConstrPsi}\\
                 &\;\;\; \gamma(0) = \mu, ~~  L \ge \int_0^1 \gamma(t) \de t\,.\nonumber
\end{align}
We observe that, rather than attempting to describe a global minimizer of this variational problem, the correct behavior is obtained by
finding a local minimizer $\gamma= \gamma_{\mu}$. Namely, we look for a local minimum of this
constrained problem with $\gamma_{\mu}$ strictly increasing on an interval $[t_0(\mu),1]$,
which is stable with respect to variations of $\gamma$ in an interval $[t_0(\mu)-\eps,1]$.
In particular, we will consider cases in which this solution is 1RSB or FRSB-1G in an interval around $\mu_{\opt}$.
We denote by
$\Psi(\mu)$ the  value of $\Par(\gamma,L)$ at this local minimizer.

Following \cite{monasson1995structural}, the function $\Psi(\mu)$ contains information about the
exponential growth rate of the number of
pure states in a system with 1RSB (scenario 1 above), and the number of ancestor states
at level $t_0$ in a system with FRSB-1G (scenario 2 above). We summarize this connection next.

Given a pure state or an ancestor state $\alpha$
(and the associated set of configurations $\cS_{\alpha}$), we let $e_{\alpha}:=\min_{\bsigma\in\cS_{\alpha}}H(\bsigma)/N$
denote the corresponding minimum energy density. The number of states (or ancestor states) at energy density $e$
is believed to grow as $\exp\{N\Sigma(e)+o(N)\}$. A parametric expression for the `complexity function' $\Sigma$ is given by
\begin{align}
  \Sigma(e) & = \mu\Psi(\mu)-\mu e\, ,\,\;\;\;  e = \frac{\partial \phantom{\mu}}{\partial\mu}\big[\mu\Phi(\mu)\big]\, .
\end{align}
We have $\mu_{\opt}: = \arg\min \Psi(\mu)$, and define $\mu_{\th}:= \arg\min \partial_{\mu} [\mu\Psi(\mu)]$: it is expected that
$\mu\mapsto \Psi(\mu)$ is decreasing for $\mu<\mu_{\opt}$ and increasing for $\mu>\mu_{\opt}$, and
$\mu\mapsto \partial_{\mu} [\mu\Psi(\mu)]$  is decreasing for $\mu<\mu_{\th}$ and increasing for $\mu>\mu_{\th}$.

We collect a few rigorously known facts about the variational problem \eqref{eq:SphericalVariational}, see \cite{chen2017parisi}:
\begin{enumerate}
\item Given $\Gamma:[0,1]\to \reals_{>0}$, define $\bar{g},g:[0,1]\to \reals$ via
  \begin{align}
    \bar{g}(t):=\xi'(t)-\int_{0}^t\frac{\de s}{\Gamma(s)^2}\, ,\;\;\;\;\;\;\; g(t) = \int_{t}^1\bar{g}(s)\,  \de s\, .\label{eq:gSphericalDef}
  \end{align}
  We let $S_0(\Gamma):=\supp(\gamma)$ be the closure of the set of points $t$ at which $\gamma$ is strictly increasing,
  with the convention that $0\in\supp(\gamma)$ if $\gamma(0^+)>0$
  (equivalently, the set of points at which $\Gamma$ is strictly concave).
  Then $\Gamma$ solves the optimization problem \eqref{eq:SphericalVariational} if and only if $\bar{g}(1)=0$, $g(t)\ge 0$ for $t\in [0,1]$, and 
  $S_0(\Gamma)\subseteq\{t\in [0,1]:\; g(t) =0\}$.
\item If $\Gamma$ solves the variational principle \eqref{eq:SphericalVariational}
  and $(a,b)\subseteq S_0(\Gamma)$ for some $0\le a<b\le 1$, then $\Gamma(t)= 1/\sqrt{\xi''(t)}$
  for $t\in (a,b)$.
\end{enumerate}

\subsection{1RSB: $\xi(x) = \frac{1}{2}x^3+\frac{1}{2}x^4$}

\begin{figure}[t]
    \centering
    \includegraphics[width=0.45\linewidth]{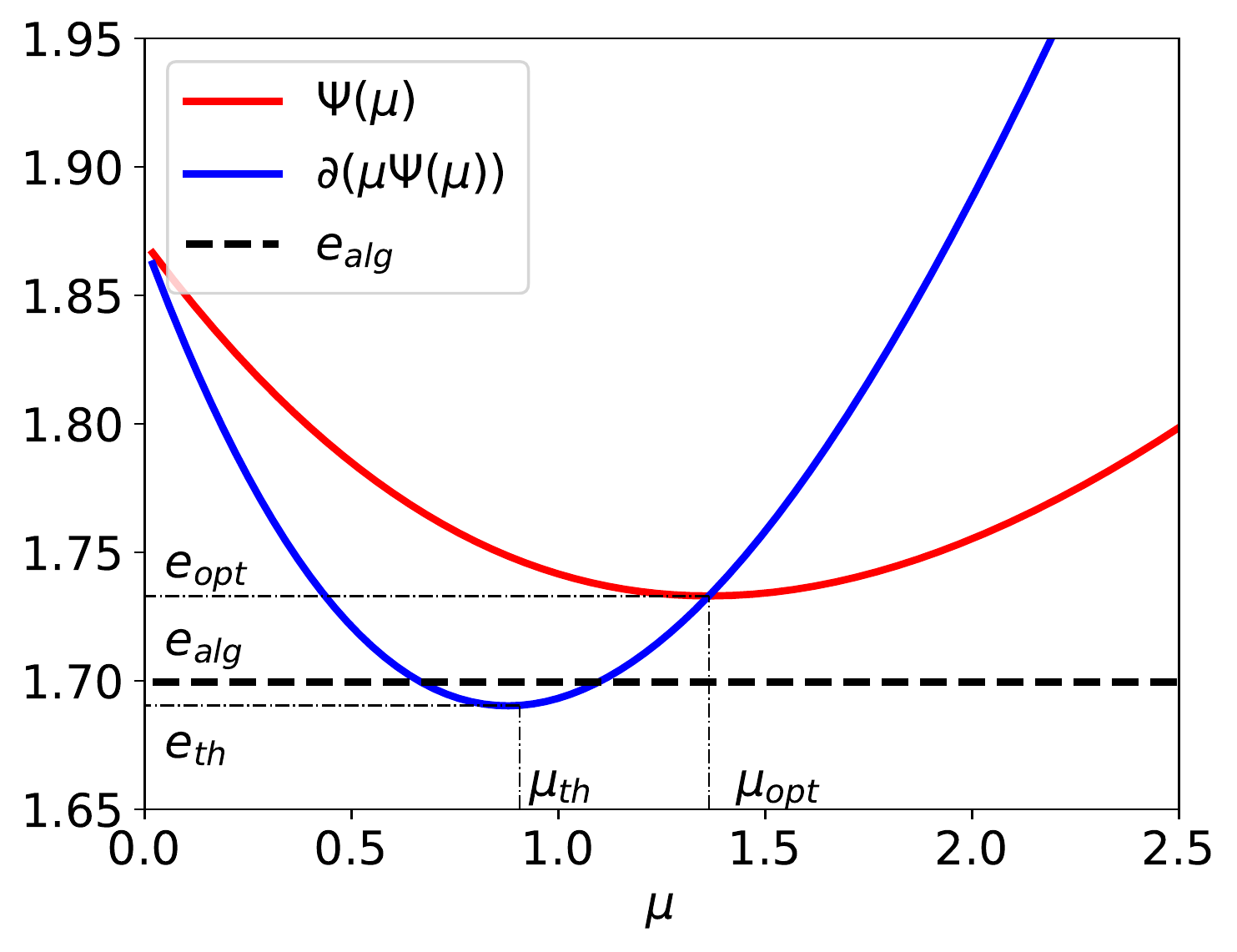}
    \includegraphics[width=0.45\linewidth]{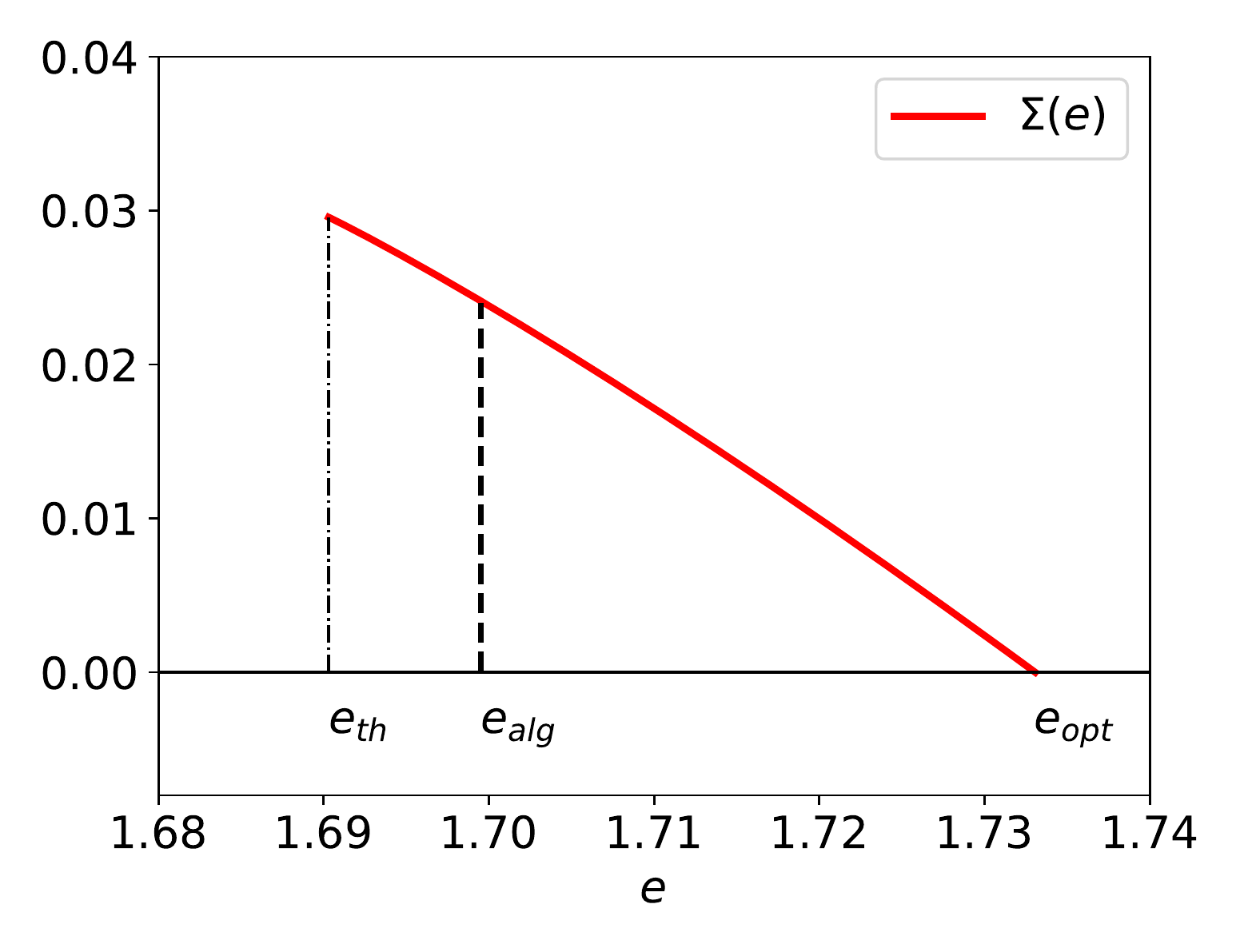}
    
    \caption{Left: The function $\Psi(\mu)$ and 
      derivative $\partial(\mu\Psi(\mu))$ for the spherical model with $\xi(x) = \frac{1}{2}x^3+\frac{1}{2}x^4$.
      The values $e_{\th}$ and $e_{\alg}$ correspond to  to the energy of threshold states and maximum energy. The dashed line to the value achieved by message passing algorithms. Right: The complexity function $\Sigma(e)$.}
    \label{fig:Cplx34}
  \end{figure}

  \begin{figure}[t]
    \centering
    \includegraphics[width=0.45\linewidth]{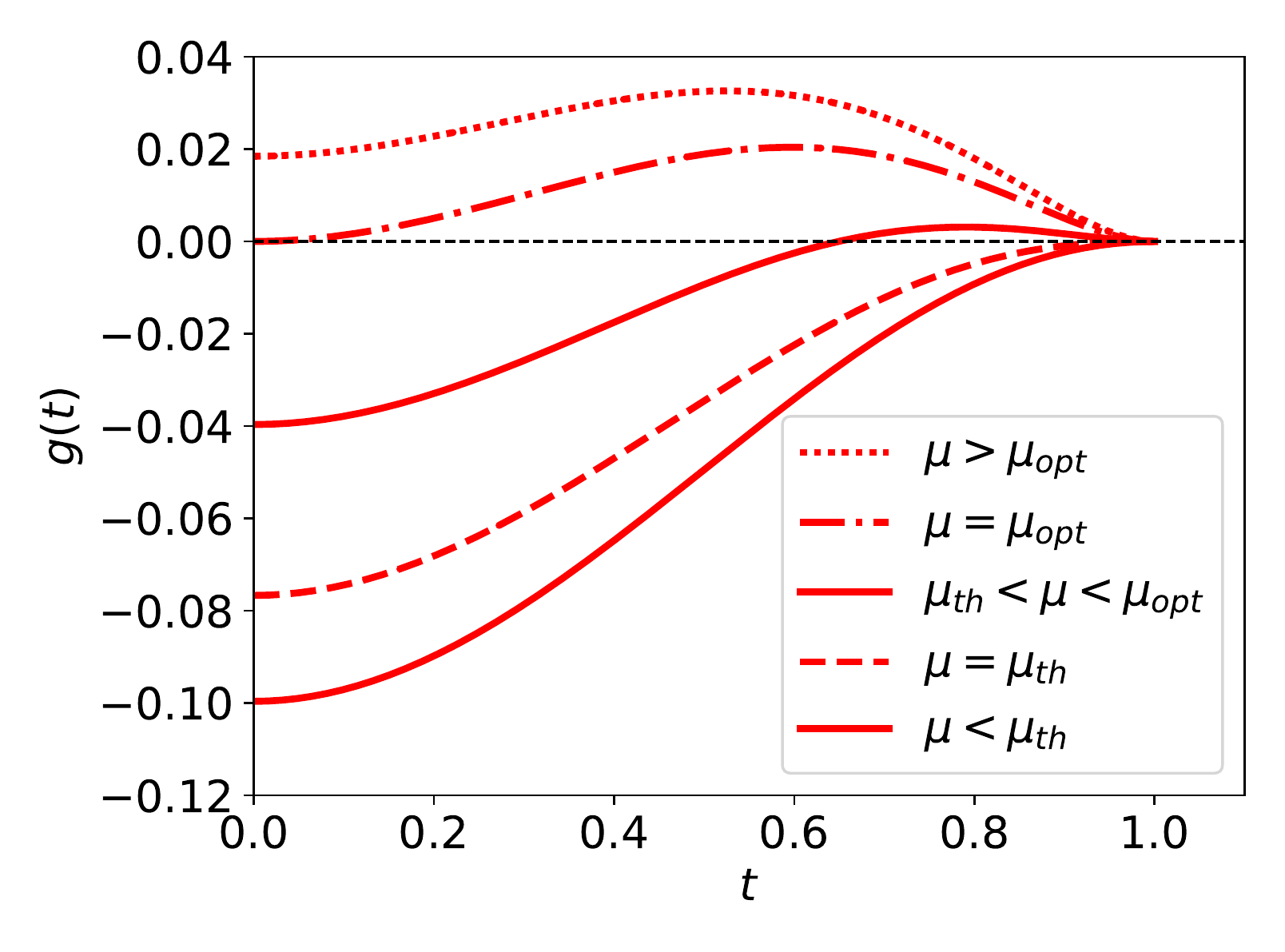}
   
    \caption{The stability criterion $g(t)$ defined in Eq.~\eqref{eq:gSphericalDef} for $\xi(x) = \frac{1}{2}x^3+\frac{1}{2}x^4$. From top to bottom
    $\mu = 0.75,  0.9072647,1.1,1.3644143,1.5$.}
    \label{fig:g34}
  \end{figure}
 
Substituting the 1RSB expression for $\gamma$ in Eq.~\eqref{eq:ParisiSpherical}, and minimizing over $L>\mu$,
we get the following well known expression
\begin{align}
  \Psi(\mu) &= \frac{1}{2}\xi'(1)L-\frac{1}{2}\mu\big(\xi'(1)-\xi(1)\big)+\frac{1}{2\mu}\log \left(\frac{L}{L-\mu}\right) \, ,\\
  & L = \frac{\mu}{2}+\sqrt{\frac{\mu^2}{4}+\frac{1}{\xi'(1)}}\, .\label{eq:Lmu}
\end{align}
It is easy to compute the derivative
\begin{align}
  \partial(\mu\Psi(\mu)) = \frac{1}{2}\xi'(1)L-\mu\big(\xi'(1)-\xi(1)\big)+\frac{1}{2(L-\mu)}\, .
\end{align}
In Figure \ref{fig:Cplx34} we plot the curves $\Psi(\mu)$, $\partial(\mu\Psi(\mu))$ for the `$3+4$' model
$\xi(x) = x^3/2+x^4/2$, as well as the resulting complexity curve.
The stability criterion $g(t)$ of Eq.~\eqref{eq:gSphericalDef} takes the form
\begin{align}
  g(t) &= \xi(1)-\xi(t)+\frac{1}{\mu L}(1-t) +\frac{1}{\mu^2}\log\Big(\frac{L-\mu}{L-\mu t}\Big)\, ,
\end{align}
where $L$ is determined by Eq.~\eqref{eq:Lmu}. In particular, expanding $g$ around $t=1$, we get:
\begin{align}
  g(t) = \frac{1}{2}\Big[\frac{1}{(L-\mu)^2}-\xi''(1)\Big](1-t)^2+O((1-t)^3)\, .
\end{align}

The maximum energy is obtained by minimizing $\Psi$ over $\mu$: the location of the minimum is
$\mu_{\opt}$, and its value is $e_{\opt} = \Psi(\mu_{\opt})$. The threshold value of $\mu_{\th}$ is obtained by solving 
$g''(1)=0$, and the corresponding energy is $e_{\th}= \partial(\mu\Psi(\mu))|_{\mu=\mu_{\th}}$.
For our running example $\xi(x) =x^3/2+x^4/2$, we obtain the following values
\begin{align}
  e_{\opt}\approx 1.733069558\, ,\;\;\;\;\;\; \mu_{\opt}\approx 1.3644143\, ,\\
  e_{\th}\approx   1.69047619047 \, ,\;\;\;\;\;\; \mu_{\th}\approx   0.9072647\, .
\end{align}
Notice that the threshold values are close, but not identical to the one corresponding to the minimum of $\partial(\mu\Psi(\mu))$,
namely $e_{\min}\approx   1.690308509$,  $\mu_{\min}\approx   0.8783100$.

The above threshold energy matches the asymptotic value achieved by gradient flow, as derived in \cite{folena2020rethinking}.
These values should be compared with the energy value achieved by the algorithm of \cite{subag2018following},
or the one studied here. This is given by Eq.~\eqref{eq:EalgSpherical}, which evaluates to
\begin{align}
  e_{\alg}\approx 1.699522254\, .
\end{align}
In other words, the present algorithms overcome the `threshold' energy.

Finally, we can compare with
an algorithm that samples an initial condition $\bsigma_0$ according to the Gibbs measure at the `mode coupling' or `dynamical
phase transition' temperature $\beta_{\rm d}$ (this is conjectured to be possible in polynomial time), and then runs gradient flow
with initialization $\bsigma_0$. The authors of \cite{folena2020rethinking} obtain the following asymptotic value for this procedure:
\begin{align}
  e_{\mbox{\rm\tiny cool}}\approx 1.696\, .
\end{align}
The algorithm studied here surpasses this value by a small but clearly non-vanishing amount.

Moving to more general models with 1RSB, the above procedure yields the following formula for the threshold energy
\begin{align}
  e_{\th} =\frac{\xi'(1)^2-\xi'(1)\xi(1)+\xi''(1)\xi(1)}{\xi'(1)\sqrt{\xi''(1)}}\,\;\;\;\;\; \mu_{\th}= \frac{\xi''(1)-\xi'(1)}{\xi'(1)\sqrt{\xi''(1)}} \, .
\end{align}
This again coincides with the asymptotic value achieved by gradient flow,
 derived in \cite{folena2020rethinking}.

\subsection{FRSB-1G:  $\xi(x) = \frac{1}{20}x^2+\frac{1}{6}x^3+\frac{1}{992} x^{32}$}

\begin{figure}[t]
    \centering
    \includegraphics[width=0.45\linewidth]{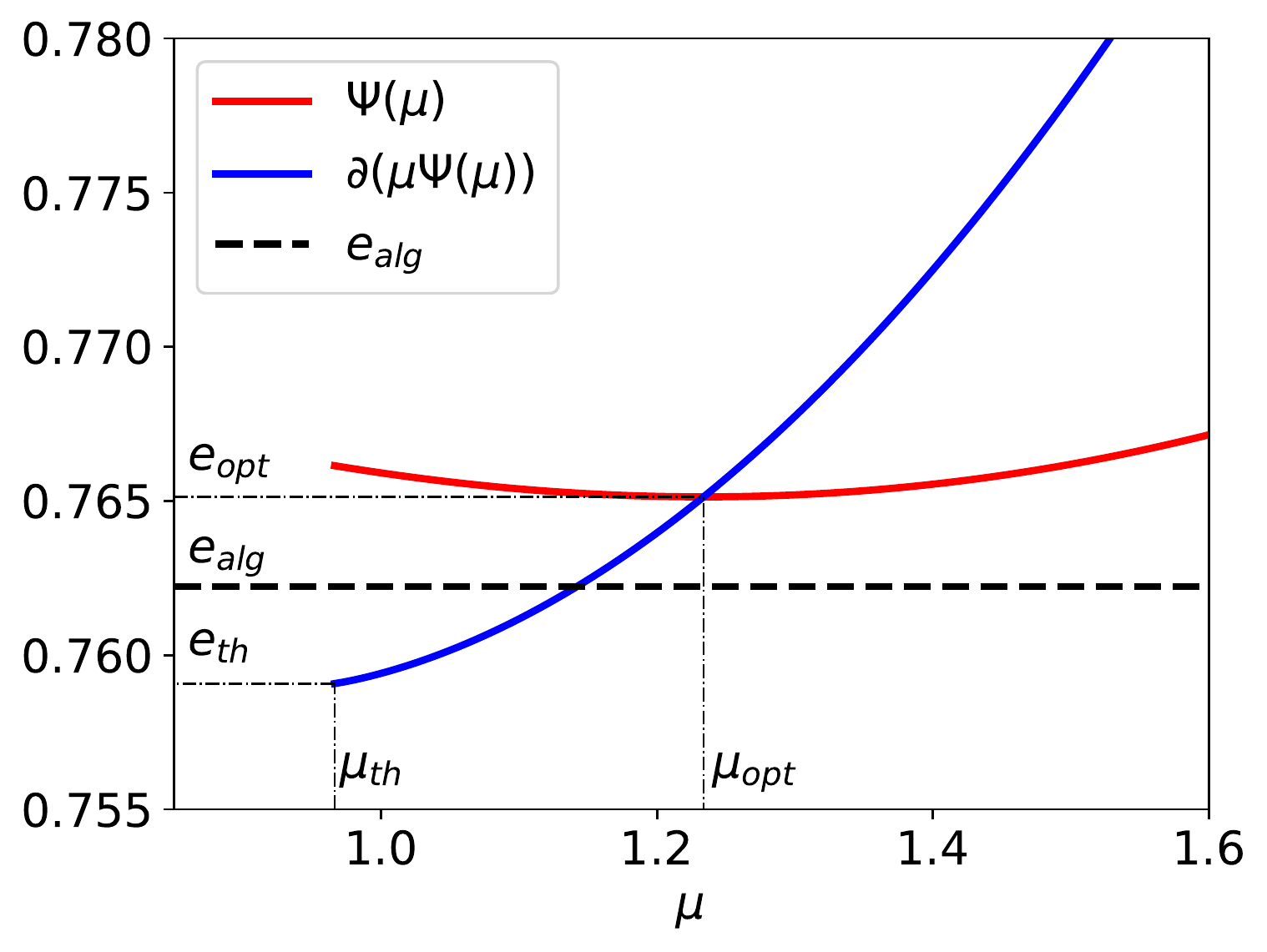}
    \includegraphics[width=0.45\linewidth]{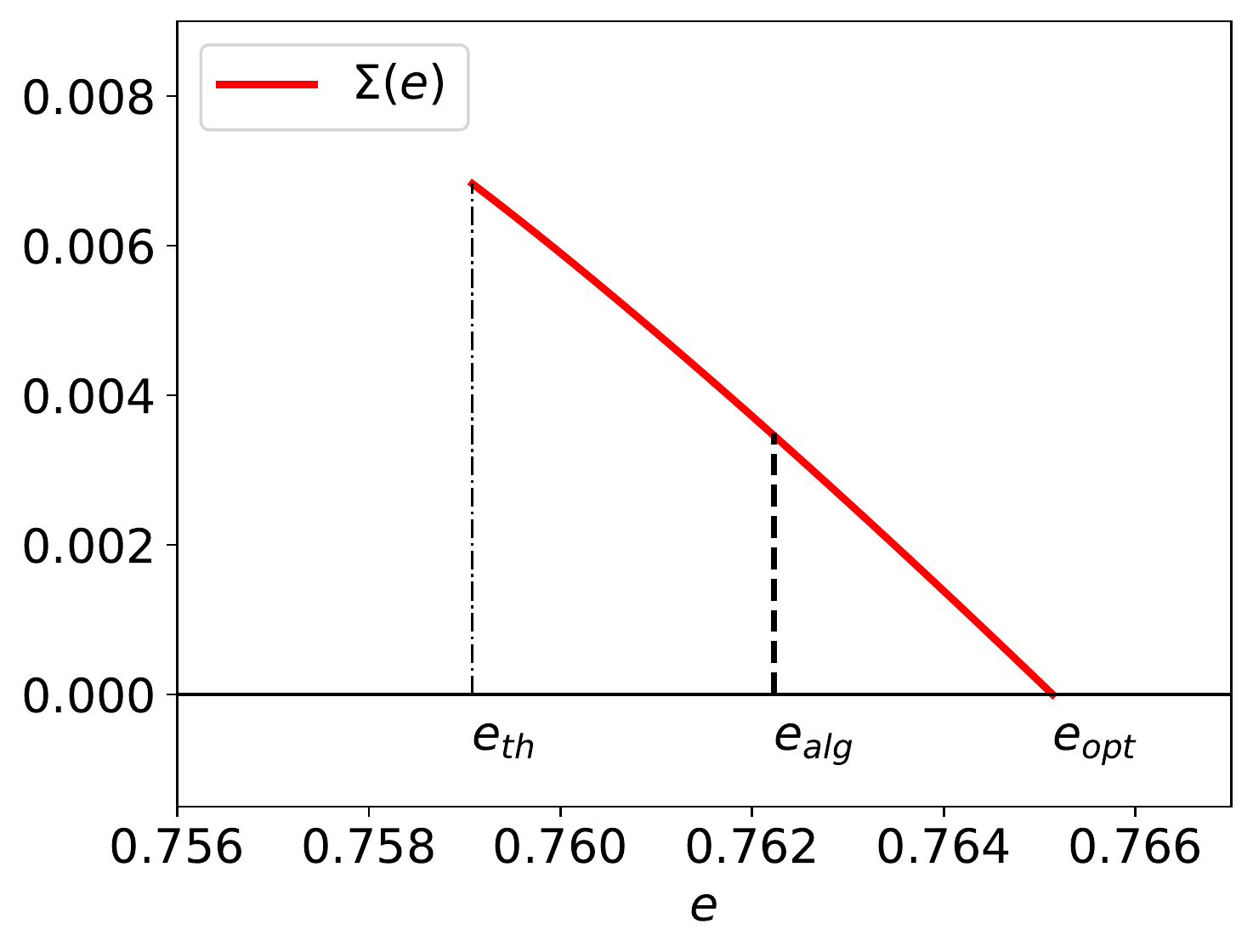}
    
    \caption{Left: The function $\Psi(\mu)$ and corresponding
      derivative for the spherical model with $\xi(x) = \frac{1}{20}x^2+\frac{1}{6}x^3+\frac{1}{992} x^{32}$.
      The two marked points correspond to the    energy of threshold states and maximum energy.
      The dashed line to the value achieved by message passing algorithms. Right: Complexity function $\Sigma(e)$.}
    \label{fig:CplxFRSB}
  \end{figure}

    \begin{figure}[t]
    \centering
    \includegraphics[width=0.45\linewidth]{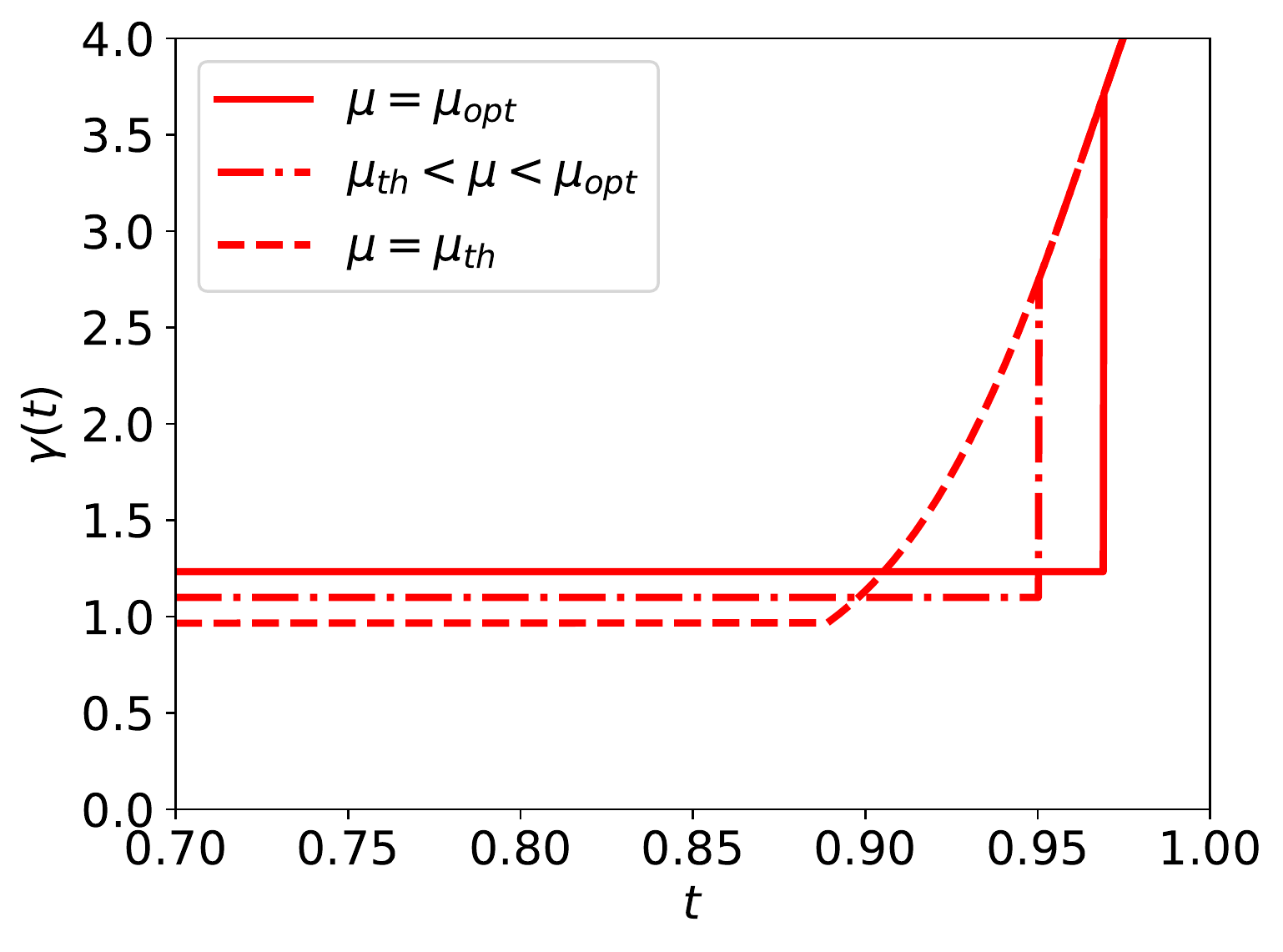}
   
    \caption{Local optimum $\gamma_{\mu}(t)$ of the constrained variational problem
     \eqref{eq:ConstrPsi} for $\xi(x) = \frac{1}{20}x^2+\frac{1}{6}x^3+\frac{1}{992} x^{32}$. From top to bottom
    $\mu \in\{ 1.2337608, 1.1, 0.96637\}$. }
    \label{fig:gFRSB}
  \end{figure}

   \begin{figure}[t]
    \centering
    \includegraphics[width=0.45\linewidth]{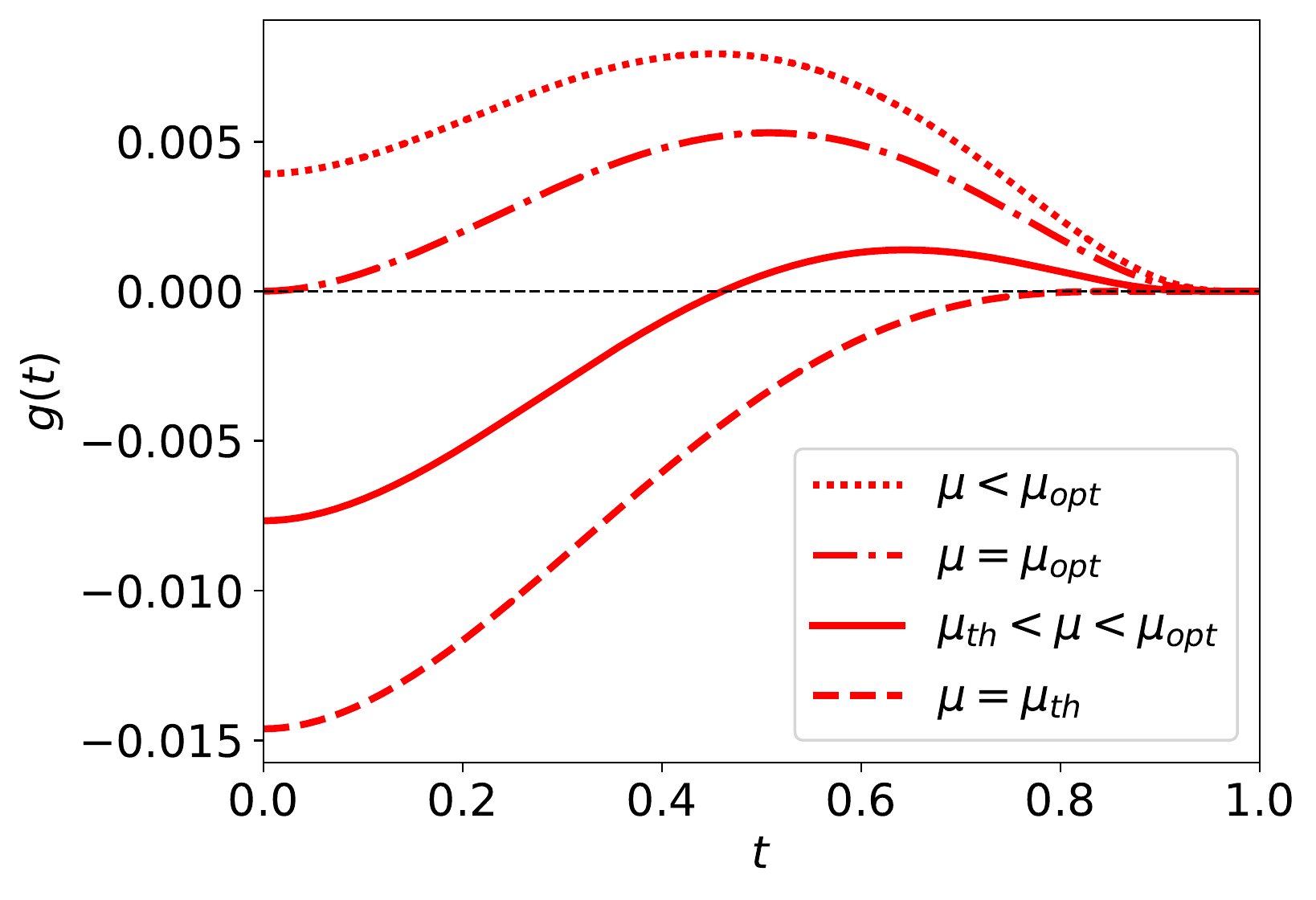}
    \includegraphics[width=0.45\linewidth]{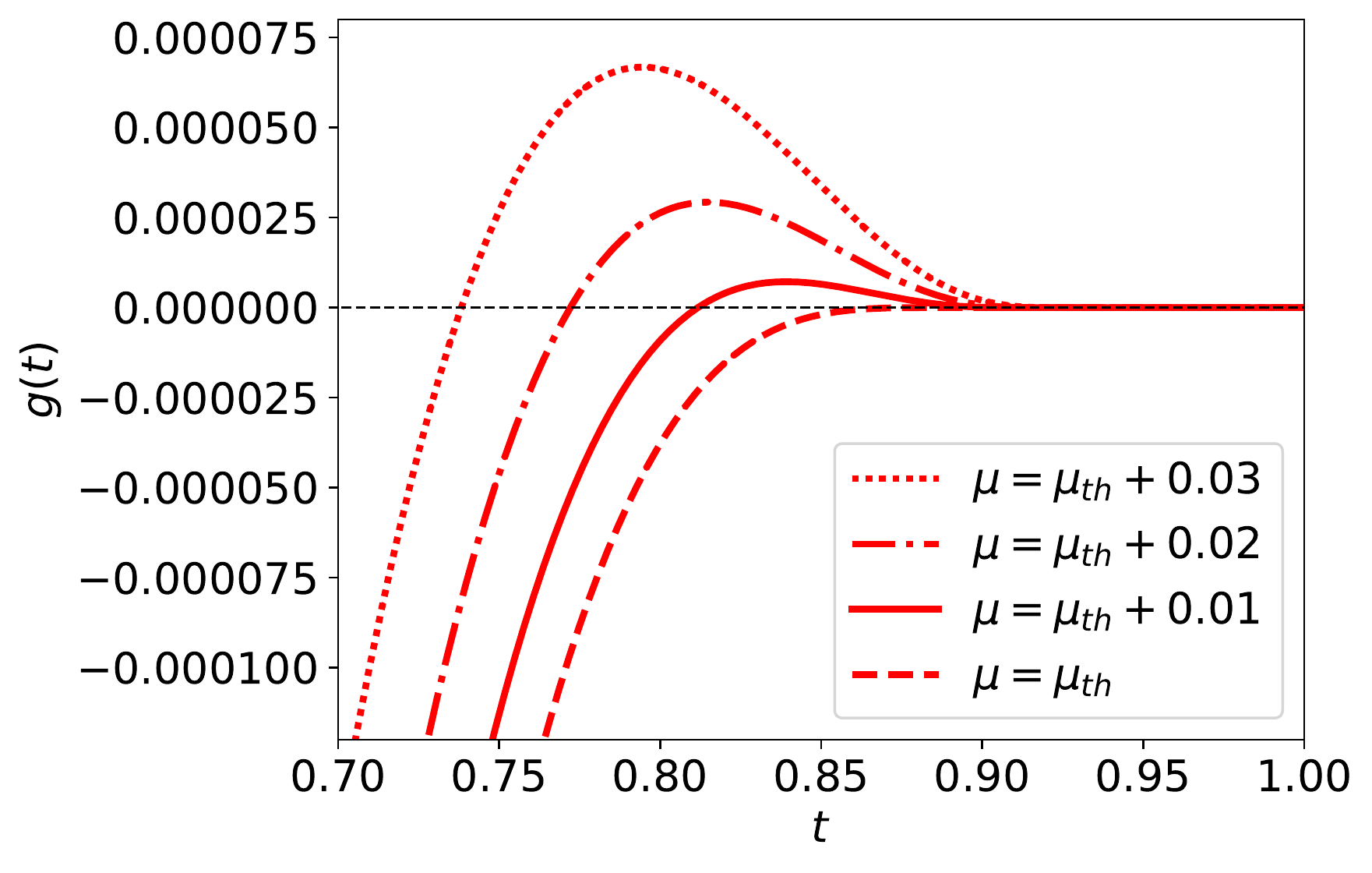}
    
    \caption{The function $g(t)$ defined in
      Eq.~\eqref{eq:gSphericalDef} for $\xi(x) =
      \frac{1}{20}x^2+\frac{1}{6}x^3+\frac{1}{992} x^{32}$. Left: from top to bottom
      $\mu = 1.3,1.2337608,1.1,0.96637$. Right: zoom around $t_*$. From top to bottom: $\mu =  0.99637,
      0.98637, 0.97637, 0.96637$.}
    \label{fig:gstab34}
  \end{figure}
 
Notice that in this case the function $t\mapsto 1/\sqrt{\xi''(t)}$ is concave over an interval $[t_s,1]$.
The value of $t_s$ is obtained by solving the equation $2\xi''(t_s)\xi^{''''}(t_s) = 3\xi'''(t_s)^2$ and in, the
present example is given by $t_s\approx 0.809049759$.
We search for a solution of the form
\begin{align}
  \Gamma(t) &= \frac{1}{\sqrt{\xi''(t_*)}} -\mu(t-t_*)\;\;\;\;\; \mbox{for $t\in [0,t_*)$}\, ,\\
  \Gamma(t)&= \frac{1}{\sqrt{\xi''(t)}}\;\;\;\;\; \mbox{for $t\in (t_*,1]$}\, .
\end{align}
Concavity of $\Gamma$ implies that $t_*\in [t_s,1]$ and
\begin{align}
  \mu \le \frac{\xi'''(t_*)}{2\xi''(t_*)^{3/2}}  \, . \label{eq:MuIneq}
  \end{align}
  Notice that  the stability criterion of Eq.~\ref{eq:gSphericalDef} requires that $\bar{g}(t)=0$ for $t\in [t_*,1]$. In particular
  $\bar{g}(1) = 0$ implies the relation
  \begin{align}
    \mu = \mu(t_*) : = \frac{\sqrt{\xi''(t_*)}}{\xi'(t_*)}-\frac{1}{t_*\sqrt{\xi''(t_*)}}\, . \label{eq:MuOfT}
  \end{align}
  Comparing this with Eq.~\eqref{eq:MuIneq}, we obtain that $t_*\ge t_{\th}$ where $t_{\th}$ is determined by the following equation
\begin{align}
  \frac{\xi'''(t_{\th})}{2\xi''(t_{\th})^{3/2}}  = \frac{\sqrt{\xi''(t_{\th})}}{\xi'(t_{\th})}-\frac{1}{t_{\th}\sqrt{\xi''(t_{\th})}}\, . \label{eq:Th}
  \end{align}
  We define the threshold value of $\mu$ by $\mu_{\th} = \mu(t_{\th})$. In Figure \ref{fig:CplxFRSB}  we plot the resulting
  function $\Psi(\mu)$ and its derivative, together with the complexity function $\Sigma(e)$. In Figure
  \ref{fig:gFRSB} we plot the corresponding function $\gamma:[0,1)\to \reals$ achieving the infimum in the variational principle.

  For our running example $\xi(x) = \frac{1}{20}x^2+\frac{1}{6}x^3+\frac{1}{992} x^{32}$, we obtain the following
  values
\begin{align}
  e_{\opt}&\approx 0.765135045\, , \;\;\;\;\;\;\;\mu_{\opt}\approx 1.2337608\, ,\\
   e_{\th}&\approx 0.759080\, , \;\;\;\;\;\;\;\mu_{\th}\approx 0.96637
\end{align}
These should be compared with the value achieved by the algorithm studied here, that is 
\begin{align}
e_{\alg} \approx 0.7622300791 \, .
\end{align}
Also in this case we observe that the algorithmic approach in this paper overcomes the threshold energy.

Finally, in order to justify the definition of the threshold $\mu$, we evaluate the stability criterion
$g(t)$ of Eq.~\ref{eq:gSphericalDef}. We obtain $\bar{g}(t) = 0$ for $t_*\le t\le 1$ and
\begin{align}
  \bar{g}(t) = \xi'(t) -\xi'(t_*) +\frac{\xi''(t_*)(t_*-t)}{1+\mu\sqrt{\xi''(t_*)} (t_*-t)}\, ,\;\;\;\;\;\; 0\le t\le t_*\, .
  \end{align}
  By our criterion for the threshold, we need to verify that $g(t)\ge 0$ for $t\in (t_*-\eps(\mu),1]$ if and only if $\mu>\mu_{\th}$.
  This is equivalent to $\bar{g}(t)\ge 0$ for $t\in (t_*-\eps(\mu),1]$ if and only if $\mu>\mu_{\th}$. Expanding
  $\bar{g}(t)$ for $t\uparrow t_*$, we obtain
\begin{align}
  \bar{g}(t) = \xi''(t_*)^{3/2}\left(\frac{\xi'''(t_*)}{2\xi''(t_*)^{3/2}}-\mu\right)(t-t_*)^2+O((t-t_*)^{3})\, .
  \end{align}
  Hence by Eqs.~\eqref{eq:MuIneq}, \eqref{eq:MuOfT}, \eqref{eq:Th},
  our definition  of $\mu_{\th}$ matches the general stability criterion.

\section{The TAP free energy is constant along the trajectory}
\label{sec:TAP_proof}
  In this Appendix we provide a short argument for the fact that the rescaled TAP free energy $\cuF_{\sTAP}(\cdot)/N$ is asymptotically constant along the trajectory $(\bm^t)_{t \in \T_{\delta} \cap [0,1]}$ of the IAMP algorithm, as $N\to \infty$ followed by $\delta \to 0$.  To this end we recall from~\cite{alaoui2020optimization} that 
  \begin{equation}\label{eq:H_N}
  \lim_{\delta \to 0^+}\, \plim_{N \to \infty} \frac{H_N(\bm^t)}{N} = \int_0^t \xi''(s) \E \big[\partial_{x}^2\Phi_{\gamma_{*}}(s,X_s)\big] \de s,
  \end{equation} 
  where $(X_t)_{t \in [0,1]}$ satisfied the SDE $\de X_t = \xi''(t) \gamma_*(t)\partial_x\Phi_{\gamma_*}(t,X_t) \de t +\sqrt{\xi''(t)} \de B_t$, with $X_0=0$, and $(B_t)_{t \in [0,1]}$ is a standard Brownian motion. 
  So it remains to find the limit of the entropy term. By state evolution and convergence of the discrete-time approximation to the SDE limit as $\delta \to 0$~\cite{alaoui2020optimization}, we have 
  \begin{equation*}
  \lim_{\delta \to 0^+}\,\plim_{N \to \infty}\frac{1}{N}\sum_{i=1}^N\Lambda_{\gamma_*}(t,m_i^t) = \E\big[\Lambda_{\gamma_*}(t,M_t)\big],
  \end{equation*}
  where $M_t = \partial_{x}\Phi_{\gamma_*}(t,X_t) = \int_0^t \sqrt{\xi''(s)} \partial_{x}^2\Phi_{\gamma_*}(s,X_s) \de B_s $. Using It\^o's formula on $ \Lambda(t,M_t)$ we have
  \begin{align*}
  \de \Lambda(t,M_t) = \Big(\partial_t\Lambda(t,M_t) + \frac{\xi''(t)}{2} \partial_m^2 \Lambda(t,M_t) \partial_{x}^2\Phi_{\gamma_*}(t,X_t)^2\Big)\de t + \partial_m \Lambda(t,M_t) \de M_t. 
\end{align*}
Exploiting the relations $ \Lambda_{\gamma_*}(t,m) := \inf_{x\in\reals}\big[\Phi_{\gamma_*}(t,x)-mx\big]$ and $M_t = \partial_{x}\Phi_{\gamma_*}(t,X_t)$ together with the strict convexity of $\Phi_{\gamma_*}(t,\cdot)$, we have 
\[\partial_t \Lambda(t,M_t) = \partial_{t}\Phi_{\gamma_*}(t,X_t), ~~\mbox{and}~~ \partial_m^2 \Lambda(t,M_t) = - \big( \partial_{x}^2\Phi_{\gamma_*}(t,X_t)\big)^{-1}.\] Therefore
\begin{align*}
\de \Lambda(t,M_t) &= \Big(\partial_{t}\Phi_{\gamma_*}(t,X_t) - \frac{\xi''(t)}{2} \partial_{x}^2\Phi_{\gamma_*}(t,X_t)\Big)\de t + \partial_m \Lambda(t,M_t) \de M_t\\
&= - \xi''(t)\Big(\frac{1}{2}\gamma_*(t)\partial_{x}\Phi_{\gamma_*}(t,X_t)^2 \de t - \partial_{x}^2\Phi_{\gamma_*}(t,X_t)\Big)\de t + \partial_m \Lambda(t,M_t) \de M_t,
\end{align*}
where we used the Parisi PDE to obtain the last line. Integrating between $0$ and $t$, 
\begin{equation}\label{eq:Lambda}
\E\big[\Lambda(t,M_t)\big] = \Phi_{\gamma_*}(0,0) - \frac{1}{2}\int_0^t \xi''(s)\gamma_*(s)\E\big[\partial_{x}\Phi_{\gamma_*}(s,X_s)^2\big]\de s - \int_0^t \xi''(s)\E\big[\partial_{x}^2\Phi_{\gamma_*}(s,X_s)\big]\de s.
\end{equation}
It was shown in~\cite{alaoui2020optimization} that the optimality of $\gamma_* \in \cuL$ as a minimizer of the Parisi formula, assuming a minimizer indeed exists, implies 
\[\E\big[\partial_{x}\Phi_{\gamma_*}(t,X_t)^2\big] = t ~~\mbox{for all}~ t \in [0,1).\]  
Putting everything together we see that the last term in~Eq.~\eqref{eq:Lambda} cancels with the energy term Eq.~\eqref{eq:H_N}, and we get for all $t \in [0,1)$,
  \begin{align*}
 \lim_{\delta \to 0^+}\,\plim_{N \to \infty}\frac{1}{N}  \cuF_{\sTAP}(\bm^t) &=  \lim_{\delta \to 0^+}\,\plim_{N \to \infty} \frac{1}{N} \Big\{ H_N(\bm^t)+\sum_{i=1}^N\Lambda_{\gamma_*}(t,m_i^t)-\frac{N}{2}\int_t^1s\xi''(s)\gamma_*(s)\,\de s\Big\}\\ 
 &=  \Phi_{\gamma_*}(0,0) - \frac{1}{2}\int_0^t \xi''(s)\gamma_*(s)s \de s  - \frac{1}{2}\int_t^1s\xi''(s)\gamma_*(s)\,\de s,\\
 &=\Phi_{\gamma_*}(0,0) - \frac{1}{2}\int_0^1 s\xi''(s)\gamma_*(s) \de s \\
 &= \Par(\gamma_*) = e_{\alg}\, .
\end{align*}
\end{document}